# On the existence of the magnetic monopole and the non-existence of the Higgs particle

Engel Roza[1]

**Summary**
In this paper the existence of the Higgs field is taken as an undeniable starting point. However, the origin of the field is challenged. Rather than ascribing the origin of it to a yet undiscovered phantom particle, the origin is ascribed directly to electromagnetic energy, in particular as magnetic charge next to electric charge of elementary point-like particles. To this end two instruments are used. The first one is the transform of the Higgs field from a functional description into a spatial description, without changing the basic properties. The other instrument is the concept of the magnetic monopole, as introduced by Dirac. The two instruments appear to fit well together. The result of all of this is that electromagnetic energy on its own is the source of all mass. It implies that the search after the Higgs particle will remain fruitless. No other equations, apart from Maxwell's Equations and Dirac's Equation are required to express the fundamentals of quantum waves and quantum fields, which makes the disputed Klein Gordon Equation obsolete. The theory reveals an algorithm to explain the ratios between the lepton masses. In that sense the theory shows a predictive element, while grosso modo, as shown, no derogation is done to the results and instruments of canonic theory.

## 1. Introduction

Since the definition of its concept by Peter Higgs in 1964 a continuous search has been made after a particular massive elementary scalar particle, known as the Higgs boson [1]. Since the unification of the weak nuclear force with electromagnetic force, by the work of Glashow, Salam and Weinberg [2,3] and final theoretical proofs by Marinus Veltman and Gerard 't Hooft [5], this particle is believed to be the corner stone of the so called Standard Model. However, in spite of efforts over more than forty years the particle has not been found. Among other reasons, the existence of such a particle is therefore not undisputed [6]. There is however firm belief that the particle will be detected soon, after full operation of the Large Hadron Collider (LHC) in the spring of 2009. The existence of a potential field of a type as produced by this particle, known as the Higgs field, is undeniable, because of many successful verifications by experiments of phenomena predicted by such fields. If the electromagnetic Maxwell laws are maintained in its classical relativistic format, there is no other way than ascribing the origin of this field to an undiscovered particle.

But what if the Maxwell laws are not maintained unimpaired? Historically a number of proposals have been put forward to generalize these laws. The most prominent ones are the generalizations

---

1. Engel Roza (1940) is a scientist retired from Philips Research Labs, Eindhoven, The Netherlands

as put forward by Paul Dirac [7] and Alexandru Proca [8]. Dirac proposed his generalization because of his wonder about an asymmetry in Maxwell's Equations. His wonder had to do with the absence of magnetic space charge in these equations as a result of a clear absence of it in experimental physics. Driven by his devotion for beauty Dirac symmetrized the equations by the hypothetical existence of a magnetic source, next to the electric source. He made clear that a verified existence of a magnetic monopole would explain the discrete nature of both electric and magnetic charges. So, a long and still continuing search began in an attempt to find experimental verification of such a magnetic monopole. Dirac himself did not urge the necessity of its existence, but never denied its existence either. Despite of efforts over more that seventy years, the magnetic monopole has not been found.

In an attempt to explain by electromagnetism the short range characteristics of nuclear forces. Proca suggested generalizing the Maxwellian Lagrangian by an extra term, proportional to mass. By doing so, Maxwellian laws would maintain its validity for zero mass and the vector potential would show an exponential decay for non-zero mass. In fact he formulated the hypothesis that a massive electromagnetic particle would exist, next to the mass-less photon. As pointed out by Yukawa [9] in 1935, this model fits with the *virtual* particle theory for bosons: a short-range force between nucleons is transferred by force-carrying particles, similarly as force exchange between charged particles is due to photons. The particles are said to be "virtual", because there is no energy available to produce the particles: they have to disappear within a time interval as imposed by the uncertainty principle. These theories appeared to be quite successful as Yukawa could predict rather accurately the characteristics of such particles, known as pi-mesons, which were verified experimentally indeed. However it appeared later that Proca's equations do not meet the so-called gauge condition of the Yang Mills Principle [10], which is believed to be a major fundament in the theory of the Standard Model. In a next section we shall elaborate on this.

In spite of these disclaimers we wish to show in this paper the feasibility of a Higgs field on the basis of generalized Maxwellian equations. We wish to show that the introduction of magnetic space charge into these equations has a short-range force resultant that matches with the characteristics of the weak nuclear force, without violating the gauge condition of the Yang Mills Principle. We shall also explain why an isolated magnetic monopole has never been found in spite of such a presumed existence of magnetic space charge. The implication of this will be that electromagnetic energy on its own, without anything else, is capable to form a fundament below the Standard Model and that no other equations apart from Maxwell's Equations and the Dirac equation are required for mathematical descriptions. There is neither need for a hypothetical elementary massive scalar particle, nor for its presumed Klein Gordon wave equation.

The paper is organized as follows. In the second section Dirac's wave equation in free space will be reviewed. This will serve as an introduction for a review of Yang Mills concept in section 3. In section 4 a view will be presented on the relationship between the concept of Lagrangian density and quantum mechanical wave equations. In section 5 an alternative description will be given for the Higgs field. Section 6 deals with wave function doublets like associated with mesons. In section 7 the origin of the Higgs field is related to magnetic monopoles. In section 8 the theory as developed in this paper is compared with the present canonic theory. Finally, in section 9 it is shown that the Higgs field as created by magnetic monopoles gives an explanation for the mass relationships between leptons (electrons, muons and tauons).

## 2. Dirac's Equation

Historically Dirac derived his equation for electrons in order to provide a relativistic wave equation as an alternative for the Klein Gordon Equation, which up to then was seen as the relativistic generalization of Schrödinger's Equation [11]. These equations are supposed to have probabilistic semantics (the so called Born interpretation), which means that the squared absolute value of the amplitude of the wave function solution represents the probability that a particle is at certain moment at a certain position. This imposes the requirement of time independency of the spatial integral of the squared absolute value of the wave function. This requirement is known as the requirement for positive definiteness. To meet this requirement, the temporal derivative in the wave equation has to be of first order. This is the case for Schrödinger's wave equation, but is not the case for the Klein-Gordon Equation. That was the basic motivation for Dirac to develop an alternative.

To keep things simple we wish to review Dirac's Equation for a single spatial dimension. This is sufficient to underline the thread of analysis in succeeding sections. With this background, generalization towards three spatial dimensions can easily be found in textbooks. Dirac has based his equation on Einstein's famous relativistic momentum relationship for moving massive particles. This relationship reads as:

$$p_0^2 + p_x^2 = -m_0^2 c^2, \tag{2-1}$$

wherein $m_0$ is the mass in rest, $c$ the velocity of light in vacuum and wherein $p_i$ are relativistic momenta. These momenta are defined as:

$$p_i = m_0 \frac{dx_i}{d\tau}, \text{ wherein } x_1 = x, x_0 = t' \text{ and } t' = jct \text{ with } j = \sqrt{-1}. \tag{2-2}$$

The momenta are expressed in proper time $\tau$, i.e. in the time frame of a co-moving observer. The normalized time coordinate $t' = jct$ is treated on par with the spatial coordinate(s). Later on the basic quantum mechanical hypothesis will be used, wherein momenta are transformed into operators on wave functions such that:

$$p_i \to \hat{p}_i \Psi \text{ with } \hat{p}_i = \frac{\tilde{h}}{j} \frac{\partial}{\partial x_i}. \tag{2-3}$$

In fact there is a slight, but not unimportant, difference between the Einsteinean relativistic energy $W$ and the energy parameter $E$ connected to the temporal moment. The Einsteinean energy $W$ is defined as:

$$W = c\sqrt{(m_0 c)^2 + p_x^2}, \tag{2-4}$$

whereas from (2-1):
$$E^2 = -p_0^2 = (m_0 c)^2 + p_x^2 .  \quad (2\text{-}5)$$

There is therefore a sign ambiguity between $W/c$ and $E$. Later on we shall come back on this.

Let us normalize (2-1) as:

$$p'^2_0 + p'^2_x + 1 = 0 \text{ with } p'_i = \frac{p_i}{m_0 c}, \ i = 0, 1 .  \quad (2\text{-}6)$$

Dirac wrote this expression as the square of a linear relationship:

$$p'^2_0 + p'^2_x + 1 = (\alpha \cdot p' + \beta) \cdot (\alpha \cdot p' + \beta) = 0 \text{ , with } \alpha(\alpha_0, \alpha_1) \text{ and } p'(p'_0, p'_x) .  \quad (2\text{-}7)$$

thereby leaving open for the moment the number type of the number $\beta$ and of the components $\alpha_0$ and $\alpha_1$ of the two-dimensional vector $\alpha$.

The elaboration of the middle term is:

$$(\alpha \cdot p' + \beta) \cdot (\alpha \cdot p' + \beta) = \left(\beta + \sum_i \alpha_i p'_i\right)\left(\beta + \sum_j \alpha_j p'_j\right)$$

$$= \beta^2 + \sum_i \beta \alpha_i p'_i + \sum_j \beta \alpha_j p'_j + \sum_i \sum_j \alpha_i \alpha_j p'_i p'_j$$

$$= \beta^2 + \sum_i (\beta \alpha_i + \alpha_i \beta) p'_i + \sum_{i \neq j} (\alpha_i \alpha_j + \alpha_j \alpha_i) p'_i p'_j + \sum_i \alpha_i^2 p'^2_i .  \quad (2\text{-}8)$$

To equate this middle term with the left hand term the following conditions should be true:

$$\alpha_i \alpha_j + \alpha_j \alpha_i = 0 \text{ if } i \neq j; \qquad \beta \alpha_i + \alpha_i \beta = 0$$

and
$$\alpha_i^2 = 1, \quad \beta^2 = 1 \quad \text{for} \quad i = 0, 1 .  \quad (2\text{-}9)$$

From these expressions it will be clear that the numbers $\alpha_i$ and $\beta$ have to be of special type. To this end Dirac invoked the use of the Pauli matrices, which are defined as:

$$\sigma_1 = \begin{bmatrix} 1 & 0 \\ 0 & -1 \end{bmatrix} \quad \sigma_2 = \begin{bmatrix} 0 & -j \\ j & 0 \end{bmatrix} \quad \sigma_3 = \begin{bmatrix} 0 & 1 \\ 1 & 0 \end{bmatrix}. \tag{2-10}$$

In addition to these also the unity matrix is required, which is defined as:

$$\sigma_0 = I = \begin{bmatrix} 1 & 0 \\ 0 & 1 \end{bmatrix}. \tag{2-11}$$

It can simply be verified that:

$$\sigma_1^2 = \sigma_2^2 = \sigma_3^2 = 1$$

$$\sigma_1\sigma_2 = j\sigma_3;\ \sigma_3\sigma_1 = j\sigma_2,\ \sigma_2\sigma_3 = j\sigma_1 \text{ and } \sigma_i\sigma_j = -\sigma_j\sigma_i \text{ for } i \neq j. \tag{2-12}$$

So, squaring of the momentum relationship, as in (2-6), can be justified if (for instance):

$$\alpha = \alpha(\sigma_3, \sigma_1) \text{ and } \beta = \sigma_2. \tag{2-13}$$

Note: It may seem that the Pauli matrices can be assigned in an arbitrary order. However, this freedom does appear not to exist. The reason is that the Dirac decomposition is not the only condition that has to be fulfilled. There is an additional constraint, which states that:

$$\Psi_1 \Psi^*_2 + \Psi^*_1 \Psi_2 = 0. \tag{2-14a}$$

This constraint is a consequence of the requirement for positive definiteness of the probability function $Pr(x, t)$ given by:

$$Pr(x, t) = \Psi_1 \Psi^*_1 + \Psi_1 \Psi^*_2 + \Psi^*_1 \Psi_2 + \Psi_2 \Psi^*_2 = |\Psi_1|^2 + |\Psi_2|^2. \tag{2-14b}$$

As we shall see, the assignment given by (2-13) will eventually yield a solution that satisfies condition (2-14a). In most textbooks this assignment problem is usually overlooked and the problem is settled by quoting something as: "among the various possibilities we choose....".

As the impulse relationship is two-dimensional, the wave function $\Psi$ should be two-dimensional as well. Therefore $\Psi = \Psi(\Psi_1, \Psi_2)$. After transforming the impulses into operators on wave functions (see 2-3), the impulse relationship is transformed into the following two-dimensional wave equation:

$$\left[\sigma_1\right]\begin{bmatrix}\hat{p}'_0\Psi_1\\\hat{p}'_0\Psi_2\end{bmatrix}+\left[\sigma_2\right]\begin{bmatrix}\hat{p}'_x\Psi_1\\\hat{p}'_x\Psi_2\end{bmatrix}+\left[\sigma_3\right]\begin{bmatrix}\Psi_1\\\Psi_2\end{bmatrix}=0,\qquad(2\text{-}15)$$

or, with explicit expressions of the Pauli matrices:

$$\begin{bmatrix}0&1\\1&0\end{bmatrix}\begin{bmatrix}\hat{p}'_0\Psi_1\\\hat{p}'_0\Psi_2\end{bmatrix}+\begin{bmatrix}1&0\\0&-1\end{bmatrix}\begin{bmatrix}\hat{p}'_x\Psi_1\\\hat{p}'_x\Psi_2\end{bmatrix}+\begin{bmatrix}0&-\text{j}\\\text{j}&0\end{bmatrix}\begin{bmatrix}\Psi_1\\\Psi_2\end{bmatrix}=0.\qquad(2\text{-}16)$$

This reads as the following two equations:

$$\hat{p}'_0\Psi_2+\hat{p}'_x\Psi_1-\text{j}\Psi_2=0\quad\text{and}\quad\hat{p}'_0\Psi_1-\hat{p}'_x\Psi_2+\text{j}\Psi_1=0,\qquad(2\text{-}17)$$

or, after denormalization (2-4):

$$\hat{p}_0\Psi_2+\hat{p}_x\Psi_1-\text{j}m_0c\Psi_2=0\quad\text{and}\quad\hat{p}_0\Psi_1-\hat{p}_x\Psi_2+\text{j}m_0c\Psi_1=0,\qquad(2\text{-}18\text{a,b})$$

or, in matrix terms:

$$\begin{bmatrix}\hat{p}_x&\hat{p}_0-\text{j}m_0c\\\hat{p}_0+\text{j}m_0c&-\hat{p}_x\end{bmatrix}\begin{bmatrix}\Psi_1\\\Psi_2\end{bmatrix}=0.\qquad(2\text{-}19)$$

Such first order partial differential equations have harmonic wave function solutions of the type:

$$\Psi_i(x,t)=u_i\exp\left[\text{j}\left(\beta\frac{p_x}{\hbar}x-\frac{E}{\hbar}t\right)\right],\qquad(2\text{-}20)$$

with parameters $u_i, p_x, E, \beta$. The characteristics of these parameters are reflected in the chosen symbols: the dimensionality of $p_x$ is that of a momentum, the dimensionality of $E$ is that of an energy, $\beta$ is dimensionless and $u_i$ is the amplitude of the wave function.

This implies that the two components of the wave function are supposed to have a similar behavior, but that they may differ in amplitude, albeit that may show a temporal time shift. The latter is the case if a phase factor is included in $u_i$, which implies that $u_i$ might be complex.

After substitution of (2-20) into (2-19) we find:

$$\begin{bmatrix} \beta p_x & j(E/c - m_0 c) \\ j(E/c + m_0 c) & -\beta p_x \end{bmatrix} \begin{bmatrix} u_1 \\ u_2 \end{bmatrix} = 0. \quad (2\text{-}21)$$

Non-trivial solutions for $u_i$ are obtained if the determinant of the matrix is zero. This is true if:

$$\frac{E^2}{c^2} = \beta^2 p_x^2 + m_0^2 c^2. \quad (2\text{-}22)$$

Comparing this expression with (2-1) and (2-5) we conclude that the condition of a zero value for the determinant reads as:

$$\beta = \pm 1. \quad (2\text{-}23)$$

This means that the assumptions we made on the character of the wave function, as expressed by (2-20) are justified. It also means a dual outcome of the determinant expression. This dual outcome simply means that the spatial part of the wave function can be a real function instead of just a complex one. And that is what we would expect from a wave function that meets the requirements for locality. Only if the wave function is spatially real it is possible to compose a spatially confined probability package out of an ensemble of individual wave functions. So, we would have to be surprised if the outcome of the determinant expression would *not* have been dual.

Let us now calculate the ratio of the amplitude values $u_i$. It follows now straightforwardly from (2-21) that

$$u_2 = \pm j \frac{v_x}{c(w+1)} \quad \text{if} \quad u_1 = 1. \quad (2\text{-}24),$$

wherein $w = \sqrt{1 - (v_x/c)^2}$.

This means that the amplitude of the second component of the wave function is usually much smaller than the first component. In the non-relativistic limit this component is negligible. The imaginary value of the amplitude of the second component implies a ninety-degree temporal phase shift of the second component as compared with the main component. Obviously there are two possible values of the phase factor. It is therefore said that the "spin" can be positive or negative.

There is something more. By substitution of (2-23) into (2-22) and subsequent evaluation under consideration of (2-4) and (2-5) we learn that:

$$E = \pm W/c. \quad (2\text{-}25)$$

That means that Dirac's Equation is not only satisfied by a positive value of *E* but by a negative value as well. As long as no semantics are connected to the parameter *E*, this sign ambiguity does not mean anything else apart from a *frequency ambiguity* in the wave function (2-20). And, according to the Born-interpretation, frequency ambiguity has no influence on the observability of a particle.

Nevertheless this phenomenon is usually being regarded as Dirac's famous puzzle of *"negative energy"*. So, what was the reason of Dirac's wonder? The reason is a wonder about the interpretation of a negative temporal moment. Obviously two types of wave function, one with a positive frequency and another with a negative frequency can satisfy Dirac's Equation. Do these two wave equations imply two types of particles or do they just imply that one and the same particle can be represented by two different wave functions? Or even stronger: does it imply that the second wave function implies that the particle can be in two different states (apart from the spin state)? It is clear that more types of particles hypothetically can satisfy these wave equations as long as the dynamics of motion are the same. But if the particles are distinguishable it can only be done by a parameter that has no influence on the dynamics of motion.

Such a particle was identified in 1932, when Anderson discovered a particle with the same rest mass as the electron, but with positive electric charge: the *positron*. Anderson's nebula chamber experiment showed the simultaneous generation from cosmic rays of electrons and positrons in parallel paths, which deviate by magnetic fields. This suggested the conversion of zero mass particles of electromagnetic energy into dual mass particles with opposite charge. This phenomenon shows that electrons and positrons are intimately related particles with the same mass but in a different state. This state difference between electrons and positrons can adequately be attributed to the difference in sign of the relativistic temporal momentum. Therefore a positron can be regarded as an electron moving backwards in time, i.e. as an electron in a different state. As a positron and electron are generated by a zero mass particle of electromagnetic energy, they may destroy each other as well, thereby generating electromagnetic radiation. The theoretical modeling of such kind of processes is beyond the scope of classical or relativistic quantum mechanics. Therefore quantum mechanical theory has been extended towards Quantum Field Theory (QFT).

The theoretical prediction of this *antiparticle* and its later experimental verification is now seen as one of the great triumphs in the history of science. The present state of art in quantum theory on nuclear particles has revealed more of those symmetries, implying many other anti-particles (*antimatter)*.

This simplified view on Dirac's Equation is helpful to highlight some consequences, which remain usually undiscussed in most textbooks. In a Side Note added to this paper one of these issues is discussed. It deals with the question about the characteristics of a valid wave equation under the assumption that the spin component is sufficiently small to be neglected.

## 3. Yang Mills Principle

So far we have only considered wave equations in free space. In this section we wish to study wave functions of particles moving in a space under influence of external fields of forces. We shall base this study upon Dirac's Equation and we will start from some observations for time-space with a single spatial dimension. As derived above, Dirac's Equation has the solution given by:

$$\Psi_1 = u_1 \exp\left[-j\frac{E}{h}t\right]\left\{\alpha_c \cos\left[\frac{p_x}{h}x\right] + \alpha_s \sin\left[\frac{p_x}{h}x\right]\right\}$$

and: $\quad \Psi_2 = ju_2 \exp\left[-j\frac{E}{h}t\right]\left\{\alpha_c \cos\left[\frac{p_x}{h}x\right] + \alpha_s \sin\left[\frac{p_x}{h}x\right]\right\}$, $\alpha_c, \alpha_s, u_1, u_2$ real valued. (3-1)

The wave function interpretation in terms of a probability density function is:

$$Pr(x, t) = \Psi_1\Psi^*_1 + \Psi_1\Psi^*_2 + \Psi^*_1\Psi_2 + \Psi_2\Psi^*_2 = |\Psi_1|^2 + |\Psi_2|^2. \tag{3-2}$$

The canceling of the cross products is due to a particular phase relationship between $\Psi_1$ and $\Psi_2$:

$$\Psi_2 = \frac{u_2}{u_1}\exp\left[-j\frac{\pi}{2}\right]\Psi_1 \tag{3-3}$$

As (3-2) is the major property for an extension of a simplex scalar wave function towards a wave function with dual format, we might ask if other relationships apart from (3-3) would provide the same property. Let us inspect (3-1) for the purpose. Under a constant phase shift $\vartheta_0$, such that:

$$\exp\left[-j\frac{E}{h}t\right] \to \exp\left[-j\left(\frac{E}{h}t + \vartheta_0\right)\right],$$

property (3-2) remains valid. This property is known under the name *global phase invariance* of the wave function. Interestingly, property (3-2) remains valid as well if the phase shift shows a spatial dependency, i.e. if

$$\exp\left[-j\frac{E}{h}t\right] \to \exp\left[-j\left(\frac{E}{h}t + \vartheta(x)\right)\right]. \tag{3-4}$$

This is known as *local phase invariance* of the wave function. Where the global phase invariance shows up as a result of an arbitrary integration constant in the solution of Dirac's Equation in free space, local phase invariance does not. So it requires an additional influence.

Yang and Mills hypothesized (in 1954) *that under influence of a field of forces the global phase invariance of quantum mechanical wave functions is changed into local phase invariance,* similarly as Einstein's hypothesis of change of global Lorentz transform invariance into local Lorentz transform invariance under similar conditions [9]. Similarly as Einstein's Principle of Equivalence this Yang Mills Principle excels in beauty and, as will be shown below, it will enable an elegant transform of free space quantum mechanical wave equations into wave equations in fields of forces.

To investigate the feasibility we apply a generic phase rotation on the wave function, such that:

$$\Psi_0 + j\Psi_1 = \exp[-j\vartheta(x)]\{\Psi_0' + j\Psi_1'\}. \qquad (3\text{-}5)$$

Herein the $\Psi' = \Psi_0' + j\Psi_1'$ represents the global phase invariant wave function and $\Psi = \Psi_0 + j\Psi_1$ represents the presupposed local phase invariant wave function. It is supposed that both these functions obey the same wave equation if this wave equation has a suitable covariant format. To establish this format, covariant derivatives have to be found in terms of a vector field $\mathbf{A}(A_0, A_x)$, which is supposed to be the cause of the change of global phase invariance into local phase invariance.

A covariant derivative $D\Psi/\partial x_i$ has to obey the property that it transforms similarly as the argument $\Psi$ (3-5), so as:

$$\frac{D\Psi}{\partial x_i} = \exp[-j\vartheta(x)]\frac{D\Psi'}{\partial x_i} \quad \text{wherein} \quad x_0 = jct \text{ and } x_1 = x. \qquad (3\text{-}6)$$

As will be shown below, this can be obtained by defining the covariant derivative as:

$$\frac{D\Psi}{\partial x_i} = \frac{\partial \Psi}{\partial x_i} + jqA_i\Psi. \qquad (3\text{-}7)$$

If $\vartheta(x)$ were a simple constant, the spatial and temporal derivatives would have the format:

$$\frac{\partial \Psi}{\partial x_i} = \exp[-j\vartheta(x)]\frac{\partial}{\partial x_i}\Psi'. \qquad (3\text{-}8)$$

But the spatial dependence spoils this simple format for the spatial derivative into:

$$\frac{\partial \Psi}{\partial x} = \exp[-j\vartheta(x)]\frac{\partial}{\partial x}\Psi' - j\exp[-j\vartheta(x)]\Psi'\frac{\partial}{\partial x}\vartheta(x). \qquad (3\text{-}9)$$

To guarantee compatibility between conditions (3-6), (3-7) and (3-8), the field components $A_i$ have to fulfill the following condition (see appendix A):

$$qA_i = qA_i' + \frac{\partial}{\partial x_i}\vartheta(x). \tag{3-10}$$

Herein $q$ is a proportionality factor, known as coupling constant. In the case of electromagnetic fields the coupling factor is identified as electric charge. Under this condition the covariant derivative has the format as defined in (3-7). It is a so-called *gauge condition*. Note that this condition is a result of the proposed format for the covariant derivative under the covariance condition. Another format would have resulted in another gauge condition. Such a gauge condition can always be *formulated* but cannot always be *met*. It is met if the Lagrangian density of the (gauge) field is invariant under the gauge condition. The Lagrangian density is expressed by:

$$\mathcal{L}_g = -\frac{1}{16\pi}\sum_i\sum_j |F_{ij}|^2 \quad \text{wherein} \quad F_{ij} = \frac{\partial A_j}{\partial x_i} - \frac{\partial A_i}{\partial x_j}. \tag{3-11}$$

It can be easily verified that under condition (3-10):

$$F_{ij} = \frac{\partial A_j}{\partial x_i} - \frac{\partial A_i}{\partial x_j} = F_{ij}' = \frac{\partial}{\partial x_i}A_j' - \frac{\partial}{\partial x_j}A_i'. \tag{3-12}$$

The underlying reason is that the addition of scalar function on $A_x$ has no influence on the magnetic field ($\mathbf{B} = \nabla \times \mathbf{A}$). Or, formulated more principally:

*The very reason that covariant derivatives can be found in terms of the vector field $\mathbf{A}$ is the fact that the Lagrangian density of this vector field is invariant under local phase rotations.*

The relevance of the considerations above is this: the wave function of particle moving in a field forces can be found by solving its wave equation. This wave equation is a transformed version of the free field wave equation. The transformation consists out of replacing normal derivatives by covariant derivatives. The format of the covariant derivative is subject to a gauge condition on the vector potential of the field forces. The format of the gauge condition results from application of the Yang Mills hypothesis. This hypothesis supposes that the global phase invariance of the wave function of a free moving particle is changed into local phase invariance if the particle is subject to a field of forces.

In fact the Yang Mills hypothesis does not reveal anything new in the case of quantum electrodynamics (QED). It is just another formulation for the so-called *principle of minimum substitution*. This principle states that the wave equation of a charged particle moving in an electromagnetic field can be found from the equation of motion in a free field after transform of the momenta by the rule

$$p_i \rightarrow p_i - \mathrm{j}qA_i \tag{3-13}$$

and subsequent application of the basic quantum mechanical theorem:

$$\hat{p}_i \Psi \rightarrow (\hat{p}_i - \text{j} q A_i)\Psi \qquad \text{with} \qquad \hat{p}_i = \frac{\tilde{h}}{\text{j}} \frac{\partial}{\partial x_i} \quad . \tag{3-14}$$

As electromagnetic theory can be captured as a subset within the larger Yang Mills framework, other fields of forces may do as well. So, nucleon forces are candidates as well. A main difference between electromagnetic force and nuclear force is the effective range of influence: where electromagnetic forces (and gravitational forces) are long ranged, nucleon forces are short ranged. As already mentioned in the introduction, in 1935, in an attempt to bring nucleon forces within the electromagnetic framework, Proca had suggested a generalization of Maxwell's Equations by introducing a mass term, which appeared to have the desired effect. In the next section, which deals with the relationship between Lagrangian density and wave equations, we shall come back on this.

## 4. Lagrangian density and wave equations

There is some ambiguity in the concept of wave equations in quantum theory. Sometimes the semantics are probabilistic and sometimes energetic. The quantum mechanical wave equation of a particle is probabilistic: the square of the absolute value of the wave function solution is the probability to find the particle at a certain time at a certain position. It is therefore prone to confusion, although not always incorrect, to derive such a quantum mechanical wave equation from a Lagrangian density [12]. Lagrangian density is an energetic concept. At the other hand electromagnetic waves are clearly energetic and there is no objection to apply the Lagrangian density as a condensed format for their description. A force sensitive particle, such as an electron, is both *target* of an energetic field and *source* of it. As Dirac's Equation (unlike the Klein Gordon Equation) has a decent Lagrangian density [13] it is possible and common practice to assemble a composite Lagrangian density to capture both these aspects. Within the scope of this paper we prefer an alternative approach. We wish to adopt initially a dual assignment: one wave function and associated wave equation for the probabilistic (fermionic) aspect and an additional separate one for the energetic (bosonic) aspect. For the bosonic aspect the Lagrangian density concept will be used, for the fermionic aspect we wish to elaborate directly in terms of a wave equation without a Lagrangian layer above. Later on we wish to rediscuss the feasibility of a composite Lagrangian density. For further clearness we shall use the symbol $\Psi$ for fermionic wave functions and the symbol $\Phi$ for the scalar part of a vector field.

Fig. 1 is a graphical illustration of the model. At the right it is indicated that the free space Dirac's Equation is directly formulated from the mass of the particle, i.e. electron. The free space wave equation is transformed under influence of an external energetic vector field (by application of Yang Mills Principle) into the "in-field wave equation". This represents the fermionic aspect. If desired, this fermionic path can be simplified by ignoring spin. The dotted boxes in the upper right part of the figure show this.

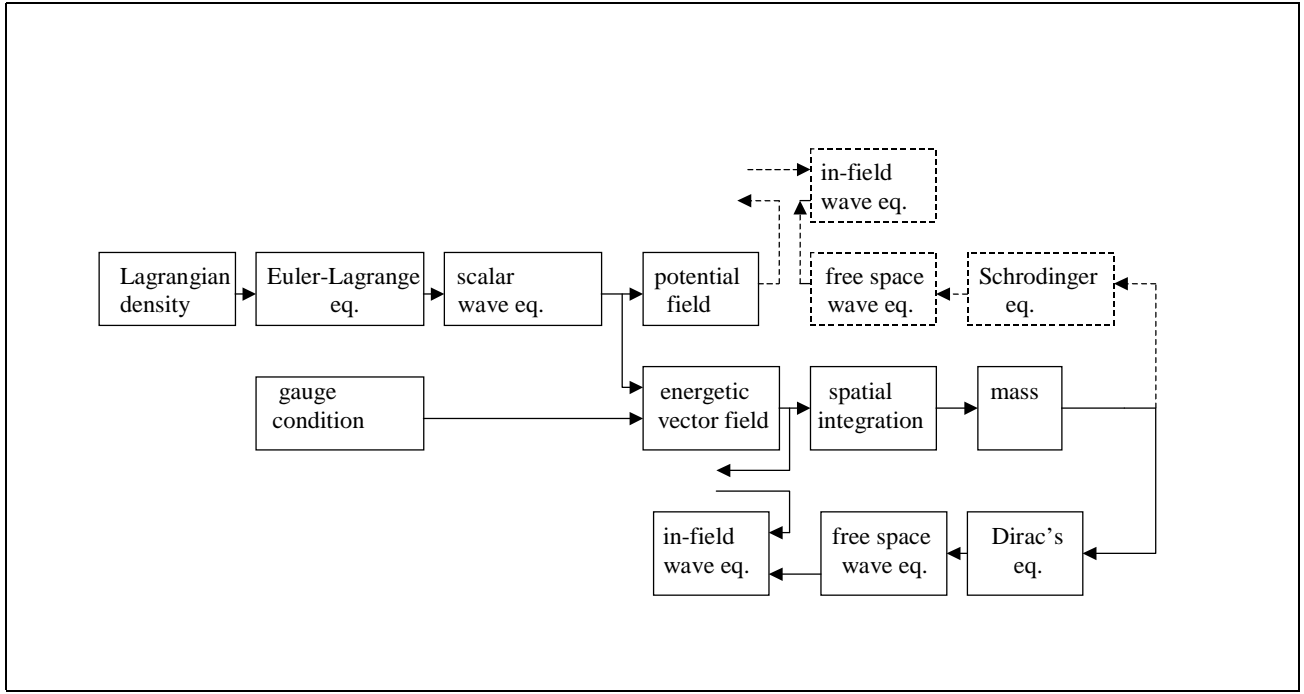

**Fig. 1: Relationship between Lagrangian density and wave equations.**

The left hand part illustrates the generation of the bosonic field, i.e. the electromagnetic field, by the particle, i.e. electron. A scalar wave equation, i.e. the potential field, is derived from the electromagnetic Lagrangian density by application of the Euler Lagrange equations. After application of the Lorenz gauge the full electromagnetic vector field is obtained, which acts as an interfering field for other particles sensitive for electromagnetic fields. In this picture self-interaction (a refinement in Quantum Field Theory) is not taken into consideration. The picture also shows the relationship between the mass of an electron and the electromagnetic energy created by it. In the case that we have to do with a point-like particle, such as an electron, the spatial integral of the components of the vector field, such as electric and magnetic field strengths, determine the mass of the particle.

Note: Usually the Lagrangian density of the electromagnetic field is expressed in terms of (3-11), covariantly written as:

$$\mathcal{L}_g = -\frac{1}{16\pi} F^{\mu\nu} F_{\mu\nu}. \tag{4-A}$$

From this Lagrangian density a wave equation is derived in terms of the (four-)vector potential **A**, which assumes a simple format under an additional condition. The format is:

$$\frac{1}{c^2}\frac{\partial^2 \mathbf{A}}{\partial t^2} - (\nabla \cdot \nabla)\mathbf{A} = \mathbf{J}. \tag{4-B}$$

The vector **J** contains the sources of the field, i.e. currents and space charge. The additional condition is known as the Lorenz gauge. It reads as:

$$\nabla \cdot \mathbf{A} + \frac{1}{c^2}\frac{\partial \Phi}{\partial t} = 0. \tag{4-C}$$

As long as we are only interested in the potential field $\Phi$, which is (apart from a proportionality constant) the very first component of the vector potential, we may work with a simplified wave equation and consequently with a simplified Lagrangian density.

The picture as shown in fig.1 has been assembled with the electron in mind. Let us forget now the electron in an attempt to generalize the picture to other type of particles, particularly with respect to the bosonic field that they may generate. Let us model this bosonic field by means of a Lagrangian density with a generic stationary part of the format:

$$\mathcal{L}_g = \frac{1}{2}(\nabla \Phi)^2 - U(\Phi). \tag{4-1}$$

From this Lagrangian the wave equation is derived via the Lagrange Euler equations, resulting into:

$$\nabla(\nabla \Phi) = \frac{d}{d\Phi}U(\Phi). \tag{4-2}$$

Wave equations are often expressed in terms of *potential functions* $V_\Phi(\Phi)$ (also called potential for short) rather than in *potential energy* $U(\Phi)$. Equation (4-2) then has the format:

$$\nabla(\nabla \Phi) = V_\Phi(\Phi)\Phi \quad \text{so that} \quad V_\Phi(\Phi)\Phi = \frac{d}{d\Phi}U(\Phi) \tag{4-3}$$

Such potential functions cannot only expressed functionally as $V_\Phi(\Phi)$ but, equivalently, spatially as well as $V(x, y, z)$.

If $U(\Phi) = 0$ and if the field is spherically symmetric, the field is Coulomb-like ($V(r) \sim 1/r$) and has therefore a long range. Short-range properties of energetic fields require particular formats for $U(\Phi)$. A characteristic example is the exponentially decaying Proca field $V(r) \sim \exp[-\lambda r]/r$, which is obtained for:

$$U(\Phi) = \frac{\lambda}{2}\Phi^2. \tag{4-4}$$

In section 3 it has been stated that the behavior of a particle sensitive for potential fields is subject to Yang Mills Principle. This principle supposes the existence of a four-vector potential $\mathbf{A}(A_0, A_x, A_y, A_z)$ wherein the scalar potential $\Phi$, possibly apart from a proportionality factor, is the very first component $A_0$. Therefore a Lagrangian density as defined by (4-1) is incomplete. The full specification requires a description for the production rules for all vector potential components and the sources of origin. In addition it requires a test on the local phase invariance on these components. So far, we have only done so for the electromagnetic field (see section 3).

Within the scope of this paper we shall not spend a further discussion on the Proca field, although it would be very instructive and not difficult to handle. The Proca field equations appear not to withstand the local invariance test. It is therefore not relevant for the thread of this paper.

Instead we wish to discuss rather extensively a potential field that has become known as the Higgs field [19,20]. Like the Proca field it is a heuristic proposal. It is specified as:

$$\mathcal{L}_g = \frac{1}{2}(\nabla\Phi)^2 - U(\Phi) \quad \text{wherein} \quad U(\Phi) = -\frac{1}{2}\mu_c^2\Phi^2 + \frac{1}{4}\lambda_c^2\Phi^4, \tag{4-5}$$

wherein $\mu_c$ and $\lambda_c$ are parameters with real values. Fig.2 shows a comparison of the functional behavior of the potential energy in the GSW-model (right) and the functional behavior of the potential energy in the Proca model (left). ,

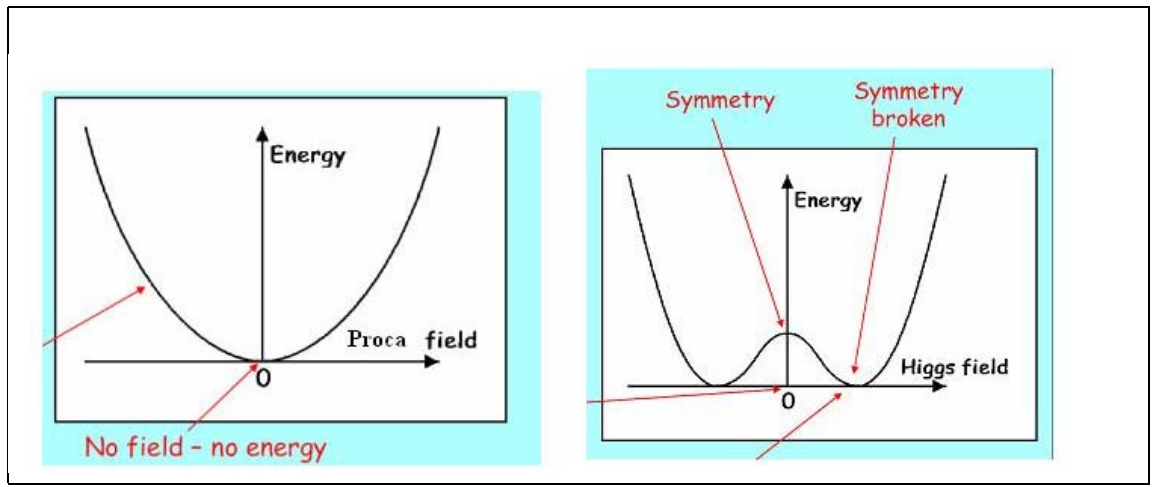

**Fig. 2: Potential energy as a function of $\Phi$.**

Although the Lagrangian density in both cases shows global phase invariance there is a symmetry shift of the minimum value of the Higgs potential. It is said that *the symmetry is broken.* As we shall see below it is this broken symmetry, which will allow local phase invariance for short-range gauge fields. Broken symmetries are not unusual in physics. Examples of broken symmetry are ferro-magnetism and super conductivity. In those cases the properties of the internal electromagnetic fields are changed under the influence of an external energetic influence, in casu under influence of temperature. The permanence of ferro-magnetism is lost above the so-called Curie temperature and super conductivity only occurs at cryogenic temperature levels. These phenomena are commonly modeled with heuristic manipulation of electromagnetic laws. In the next section we wish to present a novel analysis of the Higgs field.

## 5. The Higgs field.

We wish to consider the Higgs field under conditions of rotational symmetry. Application of the Lagrange Euler equations on the Lagrangian density as given by (4-1) gives:

$$\frac{1}{r}\frac{d^2}{dr^2}(r\Phi) = \frac{d}{d\Phi}U(\Phi) \tag{5-1}$$

sowith (4-5):
$$\frac{1}{r}\frac{d^2}{dr^2}(r\Phi) = -(\mu_c^2 - \lambda_c^2\Phi^2)\Phi. \tag{5-2}$$

This equation is difficult to solve. Rather than expanding the potential function $\Phi$ around the local minimum, like done in canonic theory, we wish to follow a different approach. This starts with profiling a tentative solution of (5-2). Let us assume that the solution of (5-2) has a format as:

$$\Phi(r) = \Phi_0\frac{\exp[-\lambda r]}{\lambda r}\left(\frac{\exp[-\lambda r]}{\lambda r} - 1\right). \tag{5-3}$$

This may seem an arbitrary guess. Let us explain the reasons for this guess. Recently Ishii [17,18] has shown a graph for the inter-nucleon potential as he with his team derived from a detailed numerical mathematical model wherein virtually all knowledge of the present state of canonic particle theory is accommodated. This is shown at the left hand part of fig.3. The right-hand part shows a curve fitting on the basis of (5-3) made by the author of this paper. The curve fits for $\lambda = 1,224 \text{fm}^{-1}$ and $V_0 = 140,55 \text{MeV}$. This curve shows the expected behavior of a combination of attractive and repulsive forces. It could well be that such a potential does not only fit at the level of the nucleon, but also at the level of a sub-nucleon. This observation will serve as the thread for our strategy to solve (5-2): Instead of solving (5-2) we adopt the spatial format of this Ishii-potential as a solution and we calculate the functional format of the right-hand part of (5-2) as a consequence of this adoption. So, in mathematical terms:

Step 1: From given $\Phi(r)$ we calculate from (5-1) a spatial expression for $dU/d\Phi$.
Step 2: We make a parametric plot of $dU/d\Phi$ versus $\Phi(r)$ (elimination of $r$).
Step 3: We apply a curve fit procedure to obtain a polynomial expression for $dU/d\Phi$ as a function of $\Phi$.

Step 4: We integrate this expression, so that $U$ as a function of $\Phi$ is obtained.

Let us normalize (5-3) as:

$$\eta(\rho) = \frac{\exp[-\rho]}{\rho}\left(\frac{\exp[-\rho]}{\rho} - 1\right) \quad \text{with} \quad \eta = \Phi/\Phi_0 \quad \text{and} \quad \rho = \lambda r. \tag{5-4}$$

:

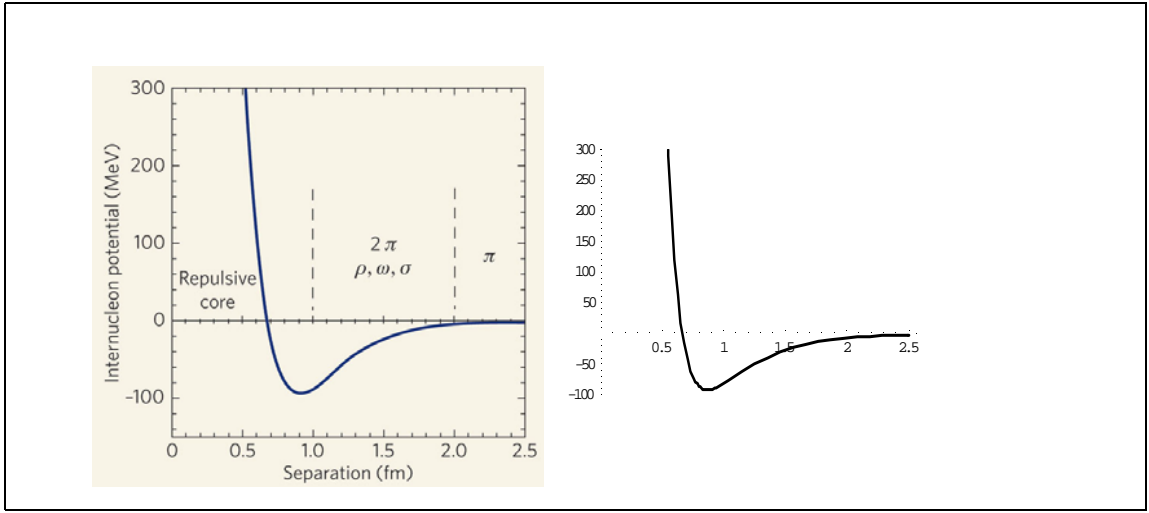

Fig. 3: **Potential function of the nucleon doublet as shown by F. Wilczek [17].**

The results of the numerical procedure are shown in fig.4. In a: $\eta(\rho)$. In b: $dU/d\eta$. In c: the parametric plot of $dU/d\eta$ vs. $\eta$ and in d: $U(\eta)$. The polynomial fit is:

$$U(\eta) = -A\eta^2 + B\eta^4 \quad \text{with} \quad A = 1,06 \quad \text{and} \quad B = 32,3 \tag{5-5}$$

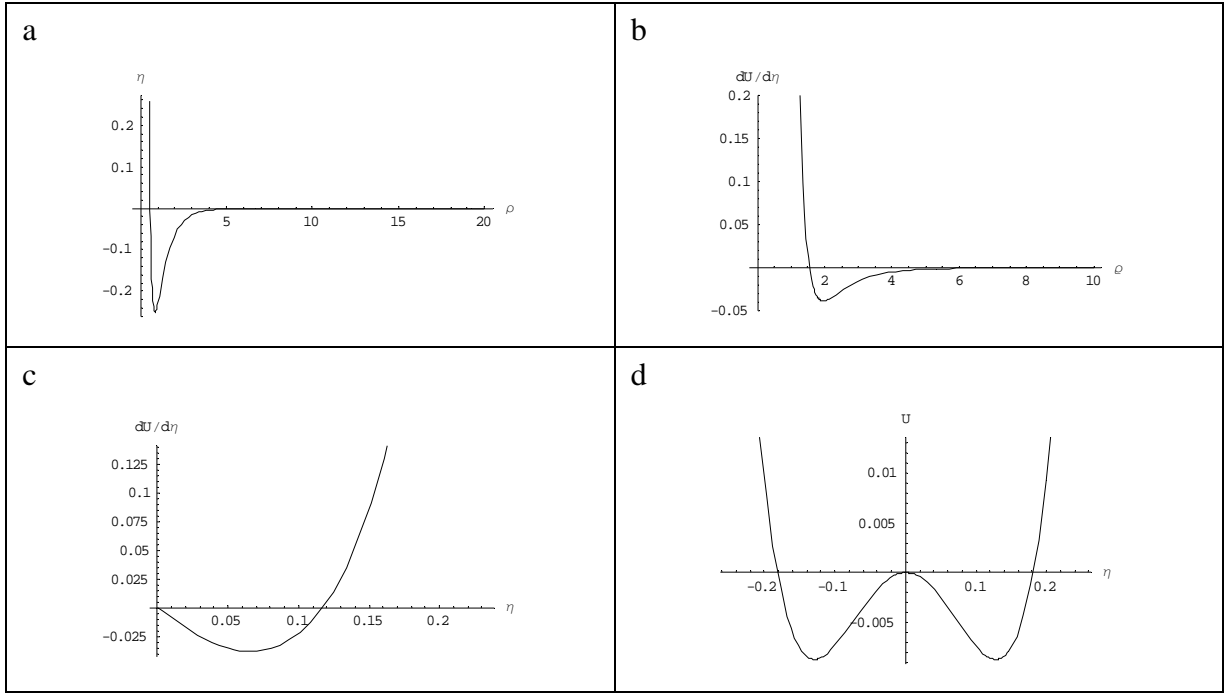

**Fig. 4: Ishii-potential vs. radius (a), functional derivative of potential energy vs. radius (b), functional derivative of potential energy vs. wave function (c), potential energy vs. wave function, i.e. Higgs field (d).**

Comparing (5-5) with (4-5), we may state that the heuristic Higgs potential matches surprisingly well with the heuristic Ishii potential. It does not mean that both formats are identical: they are just an approximation of each other. There is however no reason why one of the two is the better one. A spatial format, like the Ishii potential is, is more easily understood in its physical interpretation. Denormalization of (5-5) results in the following relationship between the parameters $\mu_c$ and $\lambda_c$ of the functional expression of the Higgs field and the parameters $\lambda$ and $\Phi_0$ of the spatial expression:

$$\frac{1}{2}\mu_c^2 = 1,06\lambda^2 \quad \text{and} \quad \frac{1}{4}\lambda_c^2 = 32,3\frac{\lambda^2}{\Phi_0^2} \tag{5-5a}$$

Some explanation is needed for the right hand part of the field equation (5-2). In classic Maxwellian electromagnetic field theory for free space the right hand part of (5-2) is zero, so that the solution of the time-independent wave equation is a Coulomb field. There is however something odd, that is to say something undefined, with the Coulomb potential. A zero right-hand part of (5-2) means that no source charge is supposed to be present. This is an oversimplification, because without a source there is no field. What is meant of course is, that there is no source charge for $r > 0$. But there must be one (with a Dirac delta-pulse shape) at $r = 0$. So, a time-independent wave equation wherein the right-hand part is spatially expressed is nothing else than Poisson's wave equation with some source for $r \geq 0$. The charge distribution is the one shown in the right upper part of fig.4 (i.e. $dU/d\eta$ ). This is rapidly decaying, so effectively zero after a very small value of *r*.

So, a *monopole* with this source gives a direct electromagnetic interpretation for the Higgs potential. Therefore, in a monopole concept there is no reason to adopt something like a *Higgs field from a hypothetical spin-less Higgs particle, which gives mass to other particles,* like it is hypothesized in the canonic theory of the Standard Model. In spite of efforts over more than forty years, such a Higgs particle has never been found. So, it is the author's belief that it does not exist, but instead that nuclear forces are carried by poin-tlike monopoles (quarks), similarly as electromagnetic forces are carried by leptons (electrons). The monopole explanation for the Higgs field gives the same unification of electromagnetic theory and theory for weak interaction as the canonic theory does, but it has the merit that no hypothetical Higgs particle is needed for this unification. In the subsequent sections of this paper this concept will be further explained.

Considering doublet structures of particles will ease the explanation. An example of a doublet is the nucleon doublet, composed by a proton and a neutron. Although these particles are not pointlike, they may be described in terms of wave functions. A proton and a neutron are subject to a combination of short-range attracting and repulsive forces such that equilibrium may occur. The attractive and repulsive forces of two nucleons will bind them together in a stationary position while the two nucleons can be vibrating. As the center of gravity will maintain a static position the system can be conceived as two individual mass-spring systems. So, there is a state of minimum energy with minimum stress on the spring. This minimum state of energy corresponds with a particular amount of spacing between the nucleons. If by slight force this initial spacing is decreased or increased and subsequently released, each of nucleons will start to perform an harmonic oscillation. This harmonic oscillation is subject to quantum mechanical laws, so each of the two systems shows ground state energy and may move to higher states of energy in quantum steps, while the nucleons keep a constant average spacing with respect to each other. This mechanism is the origin of the radiation of bosonic particles, known as pi-meseons. As mentioned in the introduction Yukawa not only predicted the existence of such particles, but was also able to predict a rather accurate estimate of their mass. Like protons and neutrons, these pi-mesons are composite particles, consisting of two quarks. So, there is good reason that doublet structures are not only apparent at the level of composite particles, but at the level of poin-tlike particles as well.

In particular doublets of particles will be considered under influence of potential fields that enable a stable configuration. As suggested above, this will bring the doublet into a status of harmonic oscillation, thereby creating conditions for absorption or radiation of bosonic particles that look like photons. This is graphically illustrated by fig.5.

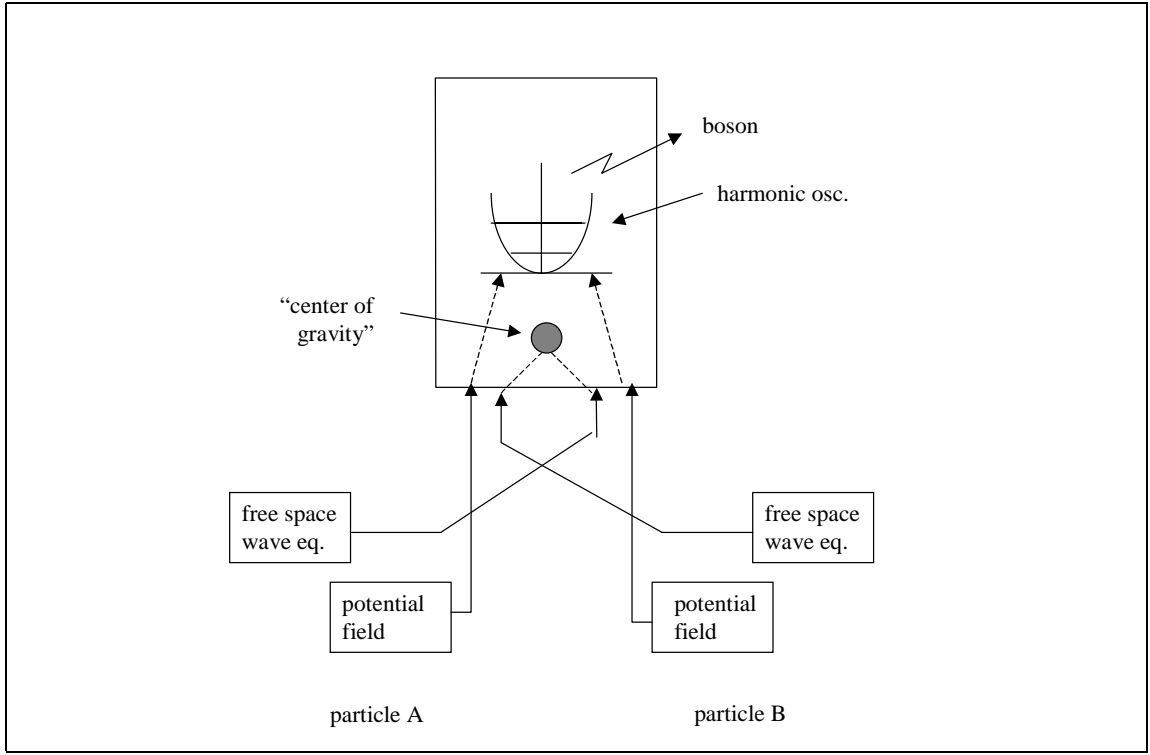

**Fig. 5: Harmonic oscillation condition of doublets.**

In the next section we wish to consider doublet structures in more detail.

## 6. The doublet.

### 6.1 The wave function doublet

Global phase invariance (and therefore local phase invariance) of a doublet is slightly different from that of a singlet in the sense that it is more than the phase invariance of two individual wave functions. Let us consider the composite $\Psi$ of the doublet, i.e.

$$\Psi = \Psi_a + \Psi_b = (\Psi_{ar} + j\Psi_{ai}) + (\Psi_{br} + j\Psi_{bi}). \tag{6-1}$$

The wave function $\Psi$ can be seen as a vector in a four-dimensional space ($\Psi_{ar}, \Psi_{ai}, \Psi_{br}, \Psi_{bi}$). What kind of phase rotation has no influence on the square of its amplitude (being the relevant parameter)? Rather than a phase rotation over a single angle, there are three possible independent angles now. Let us follow the same procedure as with (3-5) to (3-8). Local phase transformation is written as:

$$\Psi = \exp[-j\vartheta_k(x)]\Psi', \text{ with } k = 1, 2, 3. \qquad (6\text{-}2)$$

The covariant derivative has the same format as its argument, so:

$$\frac{D\Psi}{\partial x_i} = \exp[-j\vartheta_k(x)]\frac{D\Psi'}{\partial x_i}. \qquad (6\text{-}3)$$

The format of this covariant derivative is tentatively supposed to be:

$$\frac{D\Psi}{\partial x_i} = \frac{\partial \Psi}{\partial x_i} + jgB_{ki}\Psi. \qquad (6\text{-}4)$$

As compared with (3-7) the coupling factor is renamed as $g$ and the components of the now multiple vector field are renamed as $B_{ki}$.

If $\vartheta_k(x)$ were a simple constant, the spatial and temporal derivatives would have the format:

$$\frac{\partial \Psi}{\partial x_i} = \exp[-j\vartheta_k(x)]\frac{\partial}{\partial x_i}\Psi'. \qquad (6\text{-}5)$$

But the spatial dependence spoils this simple format for the spatial derivative into:

$$\frac{\partial \Psi}{\partial x} = \exp[-j\vartheta_k(x)]\frac{\partial}{\partial x}\Psi' - j\exp[-j\vartheta_k(x)]\Psi'\frac{\partial}{\partial x}\vartheta(x). \qquad (6\text{-}6)$$

To guarantee compatibility between conditions (6-2), (6-3) and (6-4), the field components $B_{ki}$ have to fulfill the following condition (see appendix A):

$$gB_{ki} = gB_{ki}' + \frac{\partial}{\partial x_i}\vartheta_k(x). \qquad (6\text{-}7)$$

This result suggests that a doublet of wave functions shows local phase invariance under influence of an assembly of three gauge fields in the case that these three fields fulfill the gauge condition. The format of the gauge condition is similar to the one for electromagnetic fields in the sense that gauge freedom should exist to add a scalar function to the spatial field vector component.

Stated otherwise: an assembly of three fields of forces is required to let two particles (wave functions) behave as a stationary assembly (doublet). The three fields are subject to an electromagnetic-like gauge condition.

Remark: the reason that three forces come forward, rather than one, is the presupposed degree of freedom of the wave function orientation in $\Psi$-space. Note that this orientation is not a spatial one. It can equally apply to a single spatial orientation of the particle assembly as to a three dimensional spatial orientation.

Experimental physics have given evidence of the existence of bosonic particles of three types indeed. At the nucleon level these particles are the mesons. Pi-mesons (pions for short) occur in two charged types: $(\pi^+)$ and $(\pi^-)$ and a neutral one $(\pi^0)$. At the sub-nucleon level three types occur as well: two charged types: $W^+$-boson and $W^-$-boson and a neutral one Z-boson.

### 6.2 The potential function of a doublet

Let us now consider two particles at a distance $d'$ apart and being subject to a force with a potential field as given by (5-3) and let these particles be responsible for the creation of this field. Let us further suppose that the two particles are aligned along the $x'$-axis and that the center of the particles is at $x' = 0$.

The result is a potential function of the type:

$$V(x) = f(x+d) + f(d-x) \quad \text{with} \quad f(x) = \frac{\exp[-2x]}{x^2} - \frac{\exp[-x]}{x},$$

wherein: $\qquad x = x'/\lambda \text{ and } d = d'/\lambda \qquad$ (6-8)

Fig. 6 shows $V(x)$ as a function of $x$, with $d$ as parameter. The function shows a clear minimum. As shown in Appendix 1, the minimum occurs for $d = 0,852$

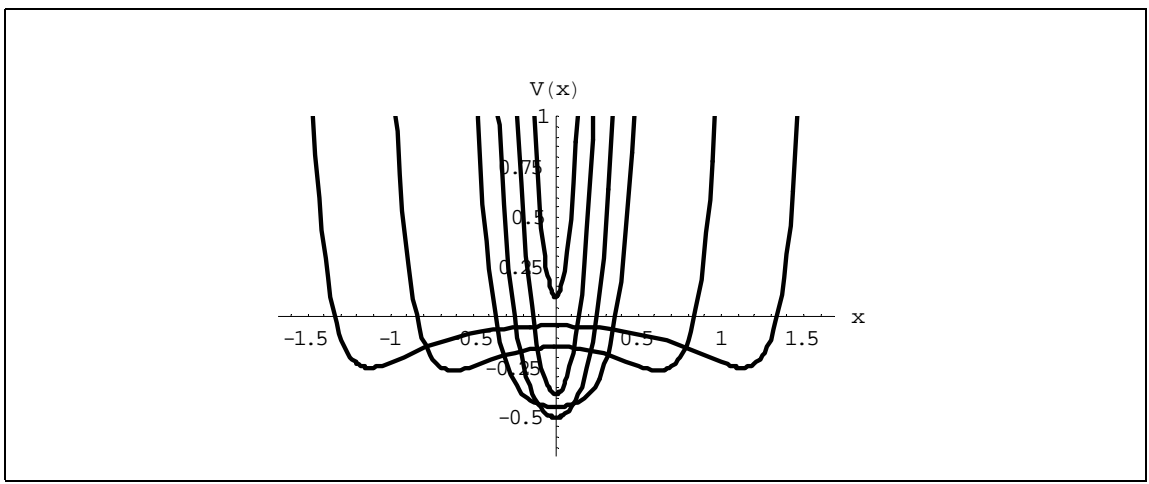

**Fig. 6: Higgs potential as a function of doublet spacing.**

Furthermore the algebraic analysis of Appendix 1 makes clear that the curve of minimum potential can be approximated as:

$$V(x) = k_0 + k_2 x^2 \text{ with } k_0 = -\frac{1}{2} \text{ and } k_2 = 2,37. \tag{6-9}$$

So, a test particle in the center of the doublet will experience a potential as given by (6-9). This potential is similar to the potential energy of a quantum mechanical harmonic oscillator. So, the test particle will behave accordingly. Of course, such a test particle is not physically present. It is represents the motion of the vibrating equilibrium of the two particles in the doublet. According to the laws of the harmonic oscillator the energy of the motion can only change in quantum steps. This corresponds with absorption or radiation of virtual particles, i.e. bosons. We may therefore conclude that this model is an elegant interpretation of the origin of such bosons in all kind of doublet structures. The most prominent among these are the nucleon doublet and the meson doublet.

## 7. The magnetic monopole

So far we have identified something as a nuclear charge having similar properties as electric charge. As sub-nucleon particles apparently are carriers of both these charge types, these charge types must simultaneously fit in electromagnetic field theory. So, where is the lacuna in the Maxwellian equations that can be used for the purpose? It was Dirac who has pointed to some asymmetry in Maxwell's Equations. This asymmetry is the absence of magnetic space charge. This absence in the equations is a consequence of the fact that no experimental physical evidence exists for the existence of such charge. Dirac disliked the asymmetry, so he reformulated Maxwell's Equations assuming existence of magnetic space charge. So Dirac's reformulation is:

$$\nabla \cdot \mathbf{E} = \rho_e \qquad \nabla \times \mathbf{E} + \frac{1}{c}\frac{\partial \mathbf{B}}{\partial t} = -\mathbf{J}_m$$

$$\nabla \cdot \mathbf{B} = \rho_m \qquad \nabla \times \mathbf{B} - \frac{1}{c}\frac{\partial \mathbf{E}}{\partial t} = \mathbf{J}_e. \tag{7-1a,b,c,d}$$

These equations are dual, i.e. they remain identical under the following transforms:

$$\mathbf{E} \to \mathbf{B} \qquad\qquad \mathbf{B} \to -\mathbf{E}$$

$$\rho_e, \mathbf{J}_e \to \rho_m, \mathbf{J}_m \qquad \rho_m, \mathbf{J}_m \to -\rho_e, -\mathbf{J}_e. \tag{7-2a,b,c,d}$$

They are generalizations of Maxwell's Equations due to the presence of $\rho_m$ and $\mathbf{J}_m$. By doing so Dirac had to find an escape route to maintain the concept of the vector potential **A**, which is defined as:

$$\mathbf{B} = \nabla \times \mathbf{A}. \tag{7-3}$$

As the rotation of a vector field is divergence free, the concept of vector potential seems to be in conflict with the presence of magnetic charge. Dirac showed that the adoption of a singularity offers a way out. The singularity has no physical impact, as it only serves to maintain the abstract mathematical vector potential construct in all space apart from the singularity. In order to have a flux from a magnetic pole while not allowing a net flux through a closed surface around the pole (zero divergence) the flux has to be brought in through a singular point on the surface. This does not mean that this mechanism has to exist in physical reality: it has no other purpose than maintaining the mathematical construct of the vector potential. This holds for two 'worlds': the normal world wherein the magnetic field is the rotation of the vector potential and the dual world wherein the electric field is the rotation of the dual vector potential.

By maintaining the vector potential concept and by assuming that a magnetic monopole would produce a similar Coulomb like field as an electric monopole does, Dirac proved that charges of electric as well as magnetic monopoles are quantized. The basic reason can be traced back to the asymmetry between two halves of the enclosed surface around the monopole: one with the singularity and the other without. Dirac has never claimed his analysis as a proof for the existence of magnetic monopoles, but he did not exclude the existence either, so that, once found, the discrete characteristics of charged particles have an explanation.

If we do not wish to adopt the existence of elementary discrete particles below the level of quarks, there is no escape from assuming a magnetic field around the quark with two types of magnetic grains: positive and negative. Therefore, unlike as the source of an electric Coulomb field, we cannot identify something like an elementary magnetic charge. The resultant of the magnetic field has a short range, which explains that an isolated magnetic monopole has never been found.

The radial magnetic field $B(r)$ can be related with the nuclear field derivative $\mathrm{d}\Phi/\mathrm{d}r$ by equating the nuclear coupling factor $g$ with the electromagnetic coupling factor. The electromagnetic coupling factor is a dimensionless quantity related with the elementary electric charge $q_e$ via:

$$q_e^2 = 4\pi\varepsilon_0 \tilde{h} c g^2. \tag{7-4}$$

In the absence of a magnetic elementary charge, the relationship of $B(r)$ with $\mathrm{d}\Phi/\mathrm{d}r$ has to be established via a relationship with the electric field.

Electromagnetic theory allows relating physical mass with energy of the electromagnetic field. A non-radiating electromagnetic cloud has energy $W_{em} = m_{em}c^2$ with

$$W_{em} = m_{me}c^2 = \iiint_V \tfrac{1}{2}\varepsilon_0 \mathbf{E} \cdot \mathbf{E}\, \mathrm{d}V + \iiint_V \tfrac{1}{2}\mu_0 \mathbf{H} \cdot \mathbf{H}\, \mathrm{d}V = \tfrac{1}{2}\varepsilon_0 \int_{r_e}^{\infty} E_r^2 4\pi r^2 \mathrm{d}r + \frac{1}{2\mu_0}\int_{r_e}^{\infty} B^2(r) 4\pi r^2 \mathrm{d}r.$$

$$\tag{7-5}$$

The shift of lower integration limit $r = 0$ to $r = r_e$ is a consequence of the *renormalization issue*. Coulomb fields ($E_r \sim 1/r^2$ would otherwise result into infinite energy. In structure based theories this problem is avoided by assuming an "empty space" for $r > r_e$ and a space with some charge distribution for $r \leq r_e$. In the formalism of QFT this issue is resolved by making a distinction between electromagnetic mass and material mass and considering the finite difference of the infinite contributions of the two at $r = 0$ as observable physical mass. As for an electron both rest mass $m_e$ and electric charge $q_e$ are observable and measurable quantities, the so-called radius of the electron $r_e$ can be established.

Although it is our aim to relate $B(r)$ with $d\Phi/dr$, we shall first relate $E(r)$ with $d\Phi/dr$ under absence of a magnetic field. We have for the force for an electric particle in an electric field:

$$F_e = q_e E_r. \tag{7-6}$$

Analogously we have for nuclear force in a nucleon field:

$$F_n = g \frac{\partial \Phi}{\partial r}. \tag{7-7}$$

Equating these two forces we get:

$$E_r = \frac{g}{q_e} \frac{\partial \Phi}{\partial r}. \tag{7-8}$$

For the energy $W_n$ of the nuclear field $W_n$, we get from (7-5):

$$W_n = \frac{1}{2} \varepsilon_0 \frac{g^2}{q_e^2} \int_0^\infty \left(\frac{\partial \Phi}{\partial r}\right)^2 4\pi r^2 dr, \tag{7-9}$$

so with (7-4):

$$W_n = \frac{1}{2\tilde{h}c} \int_0^\infty \left(\frac{\partial \Phi}{\partial r}\right)^2 r^2 dr. \tag{7-10}$$

If we now take the position that no electric energy is present in the cloud, but only magnetic energy we have from (7-10) and (7-5):

$$\frac{1}{2\tilde{h}c}\int_0^\infty \left(\frac{\partial \Phi}{\partial r}\right)^2 r^2 dr = \frac{1}{2\mu_0}\int_0^\infty B^2(r) 4\pi r^2 dr, \qquad (7\text{-}11)$$

so:
$$B(r) = \frac{g}{cq_e}\frac{d\Phi}{dr} \quad . \qquad (7\text{-}12)$$

## 8. Relationship with canonic theory

In this section we wish to compare the view as outlined above with the canonic view. First of all it has to be noted that so far our view is *Abelian,* as no a-priori field quantization is taken into account. Instead field quantization is a result from the analysis rather than a given fact. This is no conflict in views, because from this point on the *non-Abelian* instruments of field quantization can be taken up to proceed further analysis. This will enable to study particle interactions similarly as in canonic theory.

Theoretical predictions based on the hypothesized Higgs-field are so close to verifications by experiments that its existence has to be regarded to be beyond any doubt. As long as the Maxwellian laws are maintained unchanged, some artificial mechanism is required to explain the origin of the field. This mechanism might be not beyond doubt, in particular as it suggests the existence of a fancy particle subject to an additional wave equation apart from Dirac's Equation and the Maxwellian equations. Above we have shown that adaptation of Mawellian laws by hypothesizing magnetic space charge gives an adequate explanation for the origin of the Higgs field. No fancy extra particle has to be neither hypothesized, nor any other equations apart from the Maxwellian ones and Dirac's Equation.

Where canonic theory heavily relies on the elaboration of one and only Lagrangian density we have applied the Lagrangian density concept so far exclusively for the description of the energetic field, i.e. only for bosonic aspects, while canonic theory uses the Lagrangian density concept for the derivation of the probabilistic quantum mechanical wave equation as well. By doing so a composite Lagrangian density can be composed, allowing applying Feynman's methodology for field and particle interactions. As long as the quantum mechanical wave equation is Dirac's wave equation there is no conflict with the view presented in this document as Dirac's Equation can be derived from a well-defined Lagrangian density. There is no problem whatsoever to modify fig.1 accordingly. See fig. 7. In this scheme the first Lagrangian (the bosonic one) is said to perturb the second Lagrangian (the fermionic one). This scheme allows taking field quantization into account by redefining wave functions as operators on itself with a proper definition of the operation action (i.e. by changing the commutation of position and momentum by non-commutation under restriction of Heisenberg's uncertainty relationship).

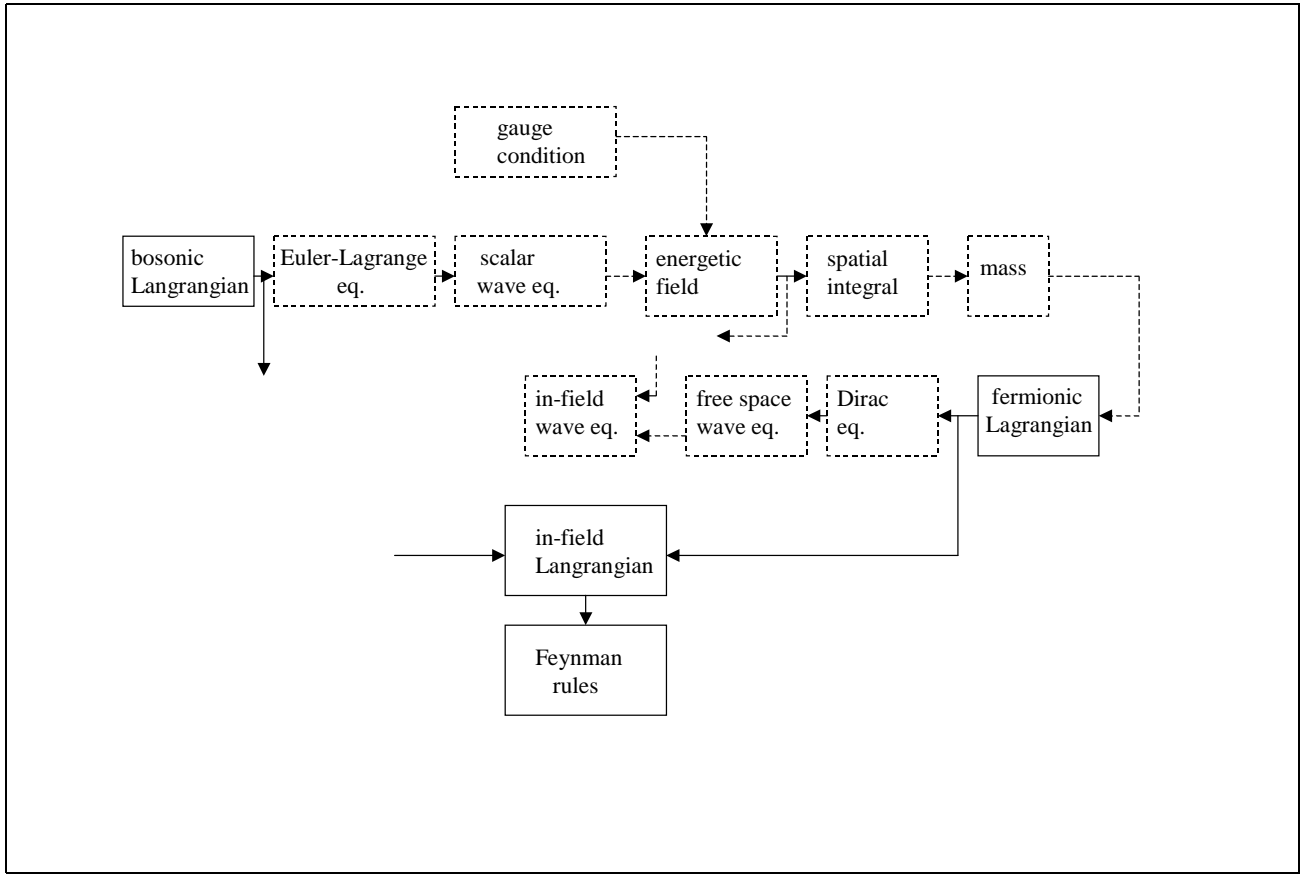

**Fig. 7: Fields, waves and Lagrangians.**

So, if it are not these considerations that make the difference, what else? Let us first consider the case of QED (Quantum Electro Dynamics). Aa single Lagrangian density, usually written as:

$$\mathcal{L} = \mathcal{L}_b + \mathcal{L}_f \tag{8-1}$$

can comprise this. $\mathcal{L}_b$ is the Lagrangian density of a (bosonic) Maxwellian field and $\mathcal{L}_f$ is the Lagrangian density of the (fermionic) Dirac field. Due to the presence of the bosonic field the fermionic wave equation is subject to Yang Mills Principle, which expresses that the global phase invariance of the free space fermionic wave function is changed into local phase invariance. This is implemented by changing common derivatives in the fermionic Lagrangian density by covariant derivatives.

Let us now apply this view to the nuclear field. In the magnetic monopole view the bosonic Maxwellian field, and therefore $\mathcal{L}_b$, is modified by the introduction of magnetic space charge while the fermionic free space wave equation is made subject to Yang Mills Principle by making $\mathcal{L}_f$ covariant. As in the case of QED, each particle has bosonic properties as well as fermionic properties. Fig. 7a and 7b shows the correspondence between the QED Lagrangian and the *WFD Lagrangian* (WFD = Weak Force Dynamics) . In canonic scientific notation the QED Lagrangian reads as:

$$\mathcal{L} = -\frac{1}{16\pi}F^{\mu\nu}F_{\mu\nu} + \overline{\Psi}(j\gamma^{\mu}\partial_{\mu} - q_e\gamma^{\mu}A_{\mu} - m)\Psi \tag{8-2}$$

Note: Greek indices are used for time-space coordinates. The coordinate $\mu, \nu = 0$ is used for the temporal coordinate. The bar in $\overline{\Psi}$ stands for the complex conjugate of the Dirac spinor and $\gamma^{\mu}$ and $\gamma_{\mu}$ are the Dirac Pauli matrices. Upper and lower indices are used according to conventions in covariant expressions. This Lagrangian is the starting point for particle interaction processes according to Feynman's methodology.

In line with the QED Lagrangian we may now formulate the WFD Lagrangian as:

$$\mathcal{L} = -\frac{1}{16\pi}F^{\mu\nu}_{(m)}F_{(m)\mu\nu} - J^{\mu}_{(m)}A_{(m)\mu} + \overline{\Psi}(j\gamma^{\mu}\partial_{\mu} - g\gamma^{\mu}A_{\mu} - m)\Psi \tag{8-3}$$

The fermionic part is the same as in the QED Lagrangian, but the bosonic part is different. This bosonic part is different from the one in the QED Lagrangian in two aspects. The electromagnetic tensor $F_{(m)\mu\nu}$ is the magnetic equivalent (i.e. the dual) of $F_{\mu\nu}$. Moreover the bosonic part is not source-less, but contains a source term $J^{\mu}_{(m)}A_{(m)\mu}$, wherein $J_{(m)\mu}$ is the magnetic current density as a consequence of the non-zero spatial span of the magnetic space charge.

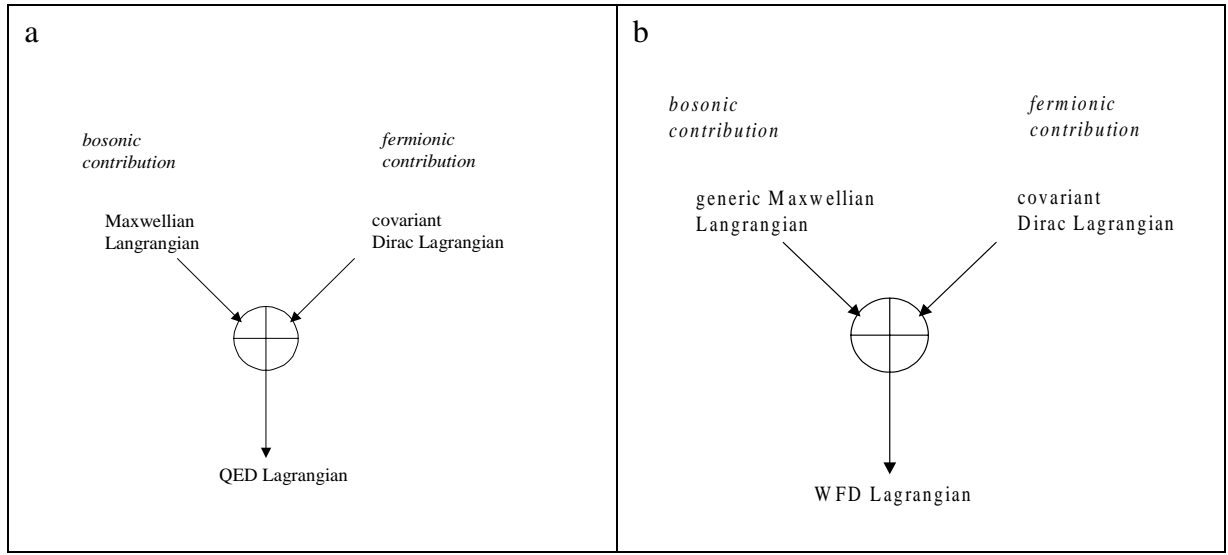

**Fig. 8: QED Lagrangian vs. WFD Lagrangian.**

As the motion of the effective mass reduces to a linear one, the four-element Dirac spinor reduces to a two element one: $\Psi(\Psi_1, \Psi_2)$. As we wish to proceed our analysis on the basis of wave equations we may apply the Euler Lagrange equations on (8-3). Equivalently, but more simply, we may derive the wave equations from the free space mechanical motion equations, application of the minimum substitution principle and the basic quantum mechanical transform of momenta into operators on wave functions. The result of, as invoked from (2-18), is:

$$(\hat{p}_0 - gA_{(m)0})\Psi_2 + (\hat{p}_x - gA_{(m)1})\Psi_1 - jcm\Psi_2 = 0$$

$$(\hat{p}_0 - gA_{(m)0})\Psi_1 - (\hat{p}_x - gA_{(m)1})\Psi_2 + jcm\Psi_1 = 0$$

wherein: $\hat{p}_i = \dfrac{\tilde{h}}{j}\dfrac{\partial}{\partial x_i}$, $x_1 = x$ and $.x_0 = jct$. (8-4)

Modeling the field as in (4-2), i.e. as a scalar field, we have $A_{(m)1} = 0$, so that from (8-4):

$$(\hat{p}_0 - gA_{(m)0})\Psi_2 + \hat{p}_x\Psi_1 - jcm\Psi_2 = 0$$

$$(\hat{p}_0 - gA_{(m)0})\Psi_1 - \hat{p}_x\Psi_2 + jcm\Psi_1 = 0.$$ (8-5)

Under suitable conditions this set of equations may have stationary solutions. Such solutions can be found from (8-5) by imposing time independency on $\Psi(\Psi_1, \Psi_2)$, which modifies the set (8-5) into:

$$\hat{p}_x\Psi_1 - (jcm + gA_{(m)0})\Psi_2 = 0 \quad \text{and} \quad \hat{p}_x\Psi_2 - (jcm - gA_{(m)0})\Psi_1 = 0.$$ (8-6)

To simplify the problem even further spin can be ignored. In the Side Note, annexed to this paper, it is derived that the spin-less limit of Dirac's Equation is:

$$j\tilde{h}\dfrac{\partial \Psi}{\partial t} + \dfrac{\tilde{h}^2}{2m_0}\dfrac{\partial^2 \Psi}{\partial x^2} - \{c^2 + V(x)\}\Psi = 0,$$ (8-7)

which reduces in the non-relativistic limit to:

$$j\tilde{h}\dfrac{\partial \Psi}{\partial t} + \dfrac{\tilde{h}^2}{2m_0}\dfrac{\partial^2 \Psi}{\partial x^2} - V(x)\Psi = 0$$ (8-8)

The time-independent form is the wave equation of a harmonic oscillator. That means that we have a straight path from the full mathematical format to a mathematical expression of a comprehensible physical system. This format will allow us to arrive at some novel insights to be discussed in next section.

# 9. Further evidence: the leptonic algorithm

What kind of experimental evidence can be brought forward to support the theory as presented above? Well, first of all we could invoke Occam's razor. It states that if two theories result in identical outcomes, the more simple of the two is the correct one. Well, inclusion of magnetic space charge in Maxwell's Equations is simpler than the adoption of an additional phantom particle. The disclaimer of course is that the subject is not dealt with in sufficient detail to claim that the experimental results following from the novel view are identical with those of the canonic theory indeed.

Secondly, we could bring forward a negative outcome of a proof: the Higgs particle has not been found so far. The disclaimer is that possibly the energies of present colliders have not been high enough to disclose its existence. Even if the LHC will not be able to disclose it, the search will probably be continued.

Thirdly, quantization of charge and consequently quantization of fields is an outcome of the theory instead of an axiom like it is in canonic theory. Also here we have a disclaimer: the magnetic monopole grain is too small to be verifiable.

If there would be any new predictive value in the theory while maintaining experimental results of the canonic theory, the theory would be superior if the predictive value can be experimentally verified. Is there some?

Let us consider the doublet potential of fig. 6 once more. Suppose that a quantum leap brings the doublet in a state of higher energy. As will be obvious from the graph, the higher state corresponds with a stronger curvature of the potential function, so a next quantum leap will correspond with an even stronger curvature. This implies that the quantum steps in the doublet's potential function will not show a constant spacing, but a progressive spacing instead. Let us try to give an estimating calculation of this spacing.

From the parabolic approximation of the potential function we have (see appendix and 6-9):

$$\frac{\Phi(r)}{\Phi_0} = k_0 + k_2(r\lambda)^2. \qquad (9\text{-}1)$$

The second term of the right hand part can be identified as the potential energy of an harmonically oscillating unknown effective doublet mass $m_p$. Relating stiffness and frequency as usual we may equate:

$$\frac{1}{2}m_p\omega^2 = \Phi_0 k_2 \lambda^2. \qquad (9\text{-}2)$$

The energetic state of the harmonic oscillator is subject to stepwise changes by the amount of $\tilde{h}\omega$. Let us identify these changes as bosonic masses $m_W$. So:

$$m_W c^2 = \tilde{h}\omega. \tag{9-3}$$

For convenience we wish to express mass in terms of energy, so we define:

$$m'_W = m_W c^2 \text{ and } m'_p \text{ accordingly.} \tag{9-4}$$

From (9-1) - (9-4) we get:

$$\Phi_0 = \frac{m'_p m'^2_W}{2(\tilde{h}c)^2 k_2 \lambda^2}. \tag{9-5}$$

Assuming near light velocity of the bosons and supposing that the boson energy is subject to the Heisenberg constraint $E\Delta t \geq \tilde{h}$, and considering that $d_0 \approx c\Delta t$, where $d_0$ is the half doublet spacing, we have for the observer in the doublet center:

$$m'_W = E = \alpha \frac{\tilde{h}c}{d_0} \tag{9-7}$$

Herein is $\alpha$ a factor to correct for the uncertainty in the Heisenberg constraint. The reason to consider half spacing rather than full spacing has to do with the modeling of the two-body system by a one-body equivalent. Note that $\alpha = 1$ corresponds with a boson wave length fit on a circle with a perimeter of $2\pi d_0$. In the one-body equivalent of a linear two-body vibrating system as we consider here it is to be expected that the spacing $d_0$ corresponds with a half wavelength fit [23]. This would make $\alpha = \pi/4$. We shall leave it as a parameter to be discussed later.

Substitution of (9-7) into (9-5) gives:

$$\Phi_0 = \frac{\alpha^2 m'_p}{2(\lambda d_0)^2 k_2}. \tag{9-9}$$

From (9-1) we note that potential function $\Phi(0)$ for generic spacing is written as:

$$\Phi(0) = \Phi_0 \{k_0(d') + k_2(d')\}, \tag{9-10}$$

wherein $d' = d\lambda$ and where, as shown in appendix B,

$$k_0 = 2\left(\frac{\exp[-2d']}{d'^2} - \frac{\exp[-d']}{d'}\right) \qquad (9\text{-}11)$$

$$\text{and } k_2 = \frac{\exp[-2d']}{d'^4}(6 + 4d'^2 + 8d') - \frac{\exp[-d']}{d'^2}\left(2 + d' + \frac{2}{d'}\right). \qquad (9\text{-}12)$$

Minimum potential occurs for $x = 0$ at $d_0' = 0,852$, so that $k_0 = -\frac{1}{2}$ and $k_2(d_0') = 2,37$.

Let us now shift the potential by an amount of $\frac{3}{2}\tilde{h}\omega = \frac{3}{2}m'_W$ and calculate the new curvature $k_2(d_a')$.

This gives the following condition:

$$\Phi_0 k_0(d_0') + \frac{3}{2}m'_W = \Phi_0 k_0(d_a') . \qquad (9\text{-}13)$$

Defining a parameter $p$ as:

$$p = \frac{m'_W}{\alpha^2 m'_p}, \qquad (9\text{-}15)$$

we get for (9-13) under consideration of (9-9):

$$k_0(d_0') + 3pd_0'^2 k_2(d_0') = k_0(d_a') . \qquad (9\text{-}14)$$

If the parameter $p$ would be known, the spacing $d_a'$ can be calculated from $d_0'$. Subsequently the calculation can be recursively repeated to calculate an even higher-energy spacing. Similarly so, the hierarchical values for the curvature $k_2$ can be established from (9-10).

In the case of the doublet archetype, i.e. the meson, all energy radiates in a boson. Therefore the vibrating mass $m_p$ and the radiative mass $m_W$ are closely related. Their semantics are different because $m_p$ is considered as non-virtual and $m_W$ as virtual. They may be at different scales of magnitude. This is due to the fact that radiative mass $m_W$ represents binding energy, only existing within the time interval of the Heisenberg uncertainty relationship. So, analysis of the harmonic oscillator has either to be made in the domain of the vibrating mass or in the domain of the radiative mass. As they show up in a ratio it does not matter which domain is chosen. We may then simply state that $m_W$ and $m_p$ are of the same of order of magnitude, which makes the parameter *p,* under consideration of the meaning of $\alpha$ :, of the order of magnitude 1.

Calculation of the first level for the hierarchical spacing above the spacing of minimum energy with $p = 1,40$ leads to $d_a' = 0,30$ and a ratio $k_2(d_a')/k_2(d_0')$ of 200. Interpretation of this result in terms of (9-2) implies a 200x increase of effective mass $m_p$ for constant $\omega$. So, if the mass at minimum energy is pure electromagnetic mass, the new electromagnetic mass is 200 times larger. If this condensates, we would have a particle with 200 times the mass of the particle in ground state. This ratio corresponds with the mass ratio of an electron and a muon. A different value for *p* would give a different ratio, so the result is somewhat 'manipulated'. Nevertheless the value $p = 1,40$ is reasonable, as will be further explained below.

In this calculation it is assumed that this first leap spans the level of minimum energy with the first escape level at a value of the equivalence of $3/2\ \tilde{h}\omega$. What about the second leap? Let us inspect the behavior of a one-dimensional quantum mechanical oscillator in terms of the Schrödinger equation, so as:

$$-\frac{\tilde{h}^2}{2m_p}\frac{d^2\Psi}{dr^2} + V(r)\Psi = E\Psi, \qquad (9\text{-}16)$$

wherein we have, as discussed above: $V(r) = \Phi_0(k_0 + k_2 r^2 \lambda^2)$.

This equation has solutions for discrete values of $(E - k_0\Phi_0)$ only, i.e. for

$$E_n = k_0\Phi_0 + \left(n + \frac{1}{2}\right)\tilde{h}\omega = \Phi_0\left\{k_0 + \left(n + \frac{1}{2}\right)\frac{m'_W}{\Phi_0}\right\}. \qquad (9\text{-}17)$$

From (9-9) and (9-14) it follows that:

$$\frac{m'_W}{\Phi_0} = 2pd_0'^2 k_2 \qquad (9\text{-}18)$$

so
$$E_n = \Phi_0\left\{k_0 + \left(n + \frac{1}{2}\right)2pd'^2 k_2\right\}, \qquad (9\text{-}19)$$

and therefore:

$$E_0 = \Phi_0\left\{k_0(d_0') + \frac{1}{2}2pd_0'^2 k_2(d_0')\right\} \text{ and } E_1 = \Phi_0\left\{k_0(d_0') + \frac{3}{2}2pd_0'^2 k_2(d_0')\right\}. \qquad (9\text{-}20)$$

At the value $E_1$ *the oscillator may jump in the ground state of a different mode* as we may equate:

$$k_0(d_0') + \frac{3}{2}2pd_0'^2 k_2(d_0') = k_0(d_1') + \frac{1}{2}2pd_1'^2 k_2(d_1'). \qquad (9\text{-}21)$$

So, if $p$ would be known, we may calculate from (9-20) the shift from $d_0'$ to $d_1'$, which establishes a narrower doublet distance for harmonic oscillation in ground mode. In fig. 9 the process is illustrated. It clarifies the meaning of the distance parameters $d_0'$, $d_a'$ and $d_1'$. The parameters $d_0'$ and $d_1'$ determine the "bottom levels" of the potential curves, while the parameters $d_a'$ and $d_b'$ determine the "take-over levels" of the modes. These levels can also be seen as "escape levels" for radiation/absorption of bosons. The $d_a'$-level is the level for the first mode (above the ground level) of the harmonic oscillation. It is at $3/2\tilde{h}\omega$ spacing from the bottom level of the lower curve. The $d_b'$-level is the level of the second mode of the harmonic oscillation. It is at $5/2\tilde{h}\omega$ spacing from the bottom level of the middle curve. The mechanism of take-over is further illustrated by the right hand part of the figure. It shows a construction of two bowls, each representing a potential curve. The vibration can be seen as a ball with some rotation energy, which can smoothly move from the lower bowl into the upper bowl. The reason why the left graph does not show a similar smooth take over is due to the mass leap to maintain a constant value for $m_W c^2 = \tilde{h}\omega$ (see 9-2).

How to interpret these results? Electromagnetic particles are subject to electric fields and magnetic fields. In classic theory magnetic fields are absent in motionless conditions. In the theory as presented above magnetic fields are present at short range in static conditions. So, a stationary doublet of electromagnetic particles may exist. In minimum energy condition these two particles are inseparable. If separated and at large distance apart they are not subject to any interaction. One constituent of the doublet after separation carries the electric field: it is the electron and its mass is determined by the electric field. The other constituent is the (electron) neutrino and carries the magnetic field. As this magnetic field is of short-range, this field does not capture a substantial amount of energy. Therefore the neutrino mass is virtually zero.

The electromagnetic doublet may be subject to external forces, which may bring the doublet in higher states of energy. These levels of energy are spaced in quantum steps. The first hierarchical level is known as a doublet of quark and antiquark, known as meson. In separated condition one of the constituents is the packet of electric energy (muon) and the other is the (muon) neutrino. The electric packet decays into smaller electric packets (electrons). The separation of the muon is the annihilation process of the quark and antiquark, which is facilitated by the virtual quantized format of the electromagnetic energy (boson).

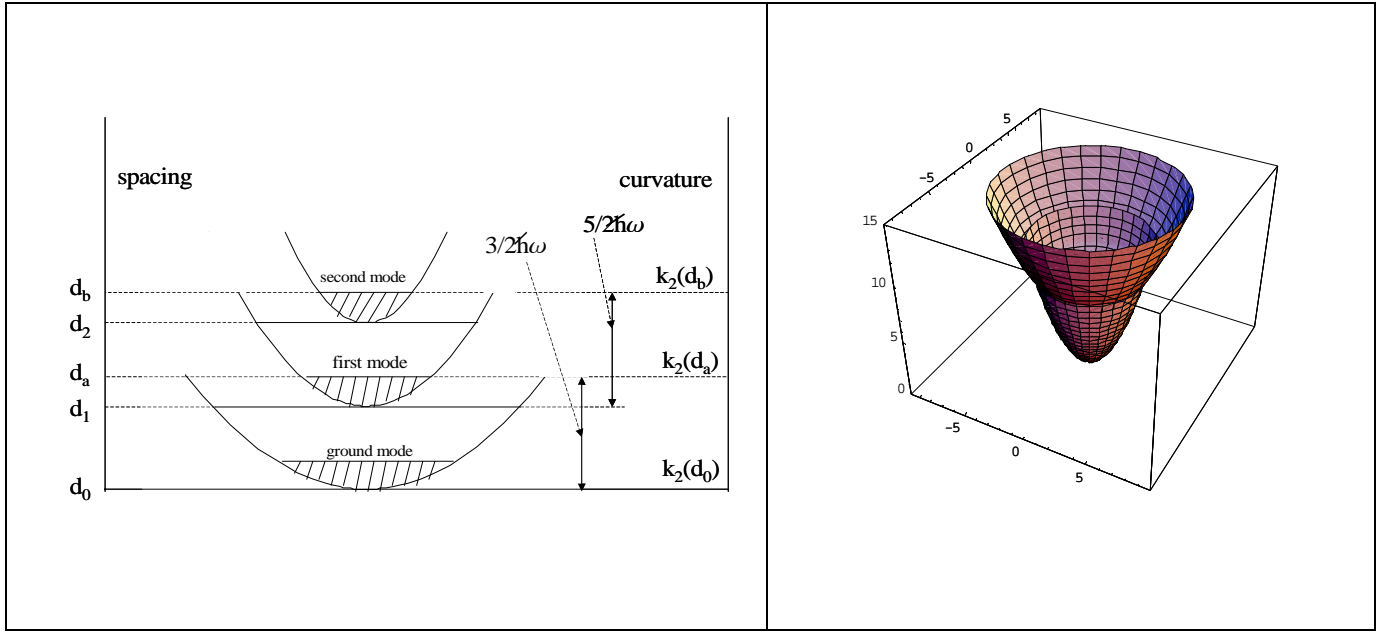

**Fig. 9: Quantum leaps of the Higgs field.**

Above we have adopted a value $p = 1,40$ to explain the mass ratio of 200 for electrons and muons. A computation of the new doublet spacing for the second mode for $p = 1,40$ yields a figure of $d_1' = 0,65$. This is different from a doublet spacing of $0,50$ which would result in ratio of 18 for tauons and muons. Unfortunately the $k_2$ ratio is rather sensitive for the parameter $p$. In fact the spacing $d_1' = 0,65$ appears to correspond with a ratio of 10,6 only. What is the reason of this discrepancy? The explanation has to do with the simple first order modeling of the potential curves by a second order function of the coordinate. This enabled us to apply the theory of a linear harmonic quantum mechanical oscillator. By expanding the potential curve to a higher order one might expect that the characteristics of the quantum leaps remain discrete, but will be subject to other spacing rules. So, it remains a challenge to refine the computation in an attempt to bring the theoretical closer to the values as found experimentally. It will require modeling the oscillator as an anharmonic quantum oscillator [21,22]. The table of fig. 10 shows our computational results for the simple harmonic model. The results are obtained, as derived above, from the following set of equations:

$$k_0(d_0') + 3p d_0'^2 k_2(d_0') = k_0(d_a')$$

$$k_0(d_0') + 3p d_0'^2 k_2(d_0') = k_0(d_1') + p d_1'^2 k_2(d_1')$$

$$k_0(d_1') + 5p d_1'^2 k_2(d_1') = k_0(d_b')$$

with (see appendix B): $k_0 = 2\left(\dfrac{\exp[-2d']}{d'^2} - \dfrac{\exp[-d']}{d'}\right)$

$$k_2 = \frac{\exp[-2d']}{d'^4}(6 + 4d'^2 + 8d') - \frac{\exp[-d']}{d'^2}\left(2 + d' + \frac{2}{d'}\right). \tag{9-22}$$

These results are fair enough to believe that our theory explains the differences and correspondences between leptons indeed, but is, unfortunately inadequate to predict the mass of the leptons beyond the tauon with some accuracy.

Nevertheless we wish to conclude that the qualititative explanation as given above reveals the existence of a leptonic algorithm. It also makes the very nature of the neutrino less mysterious.

| spacings | | | | |
|---|---|---|---|---|
| $d_0'$ | $d_1'$ | $d_a'$ | $d_2'$ | $d_b'$ |
| 0,852 | 0,66 (0,5) | 0,31 | | 0,18 |

| $k_2$-ratios | | |
|---|---|---|
| $k_2(d_a')/k_2(d_0')$ | $k_2(d_a')/k_2(d_0')$ | $k_2(d_b')/k_2(d_1')$ |
| 1 | 200 | 10,5 (18) |

| lepton masses | | |
|---|---|---|
| 0,5 MeV/c$^2$ | 100 MeV/c$^2$ | 1800 MeV/c$^2$ |

**Fig. 10: Computational results for $p = 1,40$.**

### 9.1 Anharmonic correction

We may obtain a more accurate result if the potential function as shown by (9-2) is expanded by an additional term, i.e. as:

$$\frac{\Phi(r)}{\Phi_0} = k_0 + k_2(r\lambda)^2 + k_4(r\lambda)^4. \tag{9-23}$$

Like $k_2$, the coefficient $k_4$ is a function of the normalized doublet spacing $d'$. This function can be found at the end of appendix B. The time independent part of Schrödinger's Equation now reads as:

$$-\frac{\tilde{h}^2}{2m_p}\frac{d^2\Psi}{dr^2} + \Phi_0(k_0 + k_2 r^2 \lambda^2 + k_4 r^4 \lambda^4)\Psi = E\Psi. \tag{9-24}$$

Let us rescale the variable $r$ as 
$$y = ar \tag{9-25}$$

so (9-24) can be written as:

$$-\frac{\tilde{h}^2}{4m_p k_2 \lambda^2 \Phi_0}\frac{a^4 d^2\Psi}{dy^2} + \left(\frac{y^2}{2} + \frac{k_4 \lambda^2}{2k_2}\frac{y^4}{a^2}\right)\Psi = \frac{a^2}{2\Phi_0 k_2 \lambda^2}(E - k_0\Phi_0)\Psi. \tag{9-26}$$

Let us choose the scale factor $a$ such that the coefficient of the first right hand term equals 1/2, i..e.:

$$\frac{\tilde{h}^2}{4m_p k_2 \lambda^2 \Phi_0}a^4 = \frac{1}{2}, \tag{9-27}$$

so, under definition (9-2): 
$$a^2 = \frac{m_p \omega}{\tilde{h}}. \tag{9-28}$$

Equation (9-26) can now be written as:

$$-\frac{1}{2}\frac{d^2\Psi}{dy^2} + \frac{y^2}{2} + \beta y^4 \Psi = \frac{m_p \omega}{\tilde{h}}\frac{1}{m_p \omega^2}(E - k_0\Phi_0) = \frac{1}{\tilde{h}\omega}(E - k_0\Phi_0)$$

wherein 
$$\beta = \frac{k_4 \lambda^2}{2k_2}\frac{\tilde{h}}{m_p \omega}. \tag{9-29a,b}$$

As proven in [22] this system is subject to the following relationships:

$$\frac{1}{\tilde{h}\omega}(E_n - k_0\Phi_0) = \varepsilon_n = \varepsilon_0 + n\Omega + \frac{3\beta}{2\Omega^2}n(n-1). \tag{9-30}$$

with
$$\varepsilon_0 = \frac{3}{8}\Omega + \frac{1}{8\Omega},$$

wherein $\Omega$ follows from
$$\Omega^3 - \Omega - 6\beta = 0. \tag{9-31}$$

(Note that for $\beta = 0$ the system reduces to a harmonic oscillator.)

Equation (9-31) can be simplified as follows: Let

$$m'_W = \tilde{h}\omega\Omega \quad (9\text{-}32) \text{ so under consideration of (9-7):}$$

$$\frac{\tilde{h}}{m_p\omega} = \frac{\tilde{h}^2\Omega}{m_p\tilde{h}\omega\Omega} = \left(\frac{d_0}{\alpha c}\right)^2 \frac{m'_W\Omega}{m_p} = \frac{d_0^2\Omega m'_W}{\alpha^2 m'_p} = d_0^2 p\Omega \ . \tag{9-33}$$

This equation enables to write (9-29b) as:

$$\beta = \frac{k_4}{2k_2} d'^2_0 p\Omega. \tag{9-34}$$

This result simplifies the cubic equation (9-31) into a quadratic one, because by inserting (9-34) into (9-31) we find:

$$\Omega^2 = \gamma \quad \text{wherein} \quad \gamma = 1 + 3\frac{k_4}{k_2} d'^2_0 p \ . \tag{9-35}$$

Let us now reformulate the leptonic algorithm set. What is the first escape level? From (9-28) it follows that it is a shift from the bottom by an amount of:

$$\tilde{h}\omega(\varepsilon_0 + \Omega) = m'_W\left(\frac{11}{8} + \frac{1}{8\Omega^2}\right). \tag{9-36}$$

Clearly this shift is different from the shift of $3\tilde{h}\omega/2$ as in the case of the harmonic oscillator. Consequently (9-14) has to be reformulated as:

$$k_0(d_0') + 3p_c d'^2_0 k_2(d_0')\Omega = k_0(d_a') \ ,$$

wherein
$$p_c = 2p\left(\frac{11}{8} + \frac{1}{8\Omega^2}\right)/3 \ . \tag{9-37}$$

What about the take over level? From (9-28) it follows that the ground level is defined by:

$$\tilde{h}\omega\varepsilon_0 = m'_W\left[\frac{3}{8} + \frac{1}{8\Omega^2}\right] . \tag{9-38}$$

Consequently (9-21) has to be reformulated as:

$$k_0(d_0') + 3p_c d_0'^2 k_2(d_0')\Omega = k_0(d_1') + p_{cc}\Omega d_0'^2 k_2(d_1') \quad \text{with } p_{cc} = 2p\left[\frac{3}{8} + \frac{1}{8\Omega^2}\right]. \tag{9-39}$$

Similarly as before, the equations (9-37) and (9-39) will enable to compute escape levels and take over levels. In the case of the harmonic oscillators it has been supposed that mass lepton ratios have to do with the $k_2$-ratios in order to maintain a constant value for the boson mass $m'_W$. From (9-32) we learn that this is no longer justified in the case of the anharmonic oscillator, as we have to state now:

$$\omega\Omega = \text{constant} \quad \text{so} \quad \Omega\sqrt{\frac{k_2}{m}} = \text{constant} \tag{9-40}$$

and therefore:

$$\frac{m_{mu}}{m_{el}} = \frac{k_2(d'_a)\gamma(d'_a)}{k_2(d'_0)\gamma(d'_0)} \quad \text{and} \quad \frac{m_{tau}}{m_{mu}} = \frac{k_2(d'_b)\gamma(d'_b)}{k_2(d'_a)\gamma(d'_a)} . \tag{9-41}$$

| spacing | | | | | |
|---|---|---|---|---|---|
| $d_0'$ | $d_1'$ | $d_a'$ | $d_2'$ | $d_b'$ | $d_c'$ |
| 0,852 | 0,51 | 0,16 | 0,48 | 0,08 | 0,05 |

| mass ratios | | | |
|---|---|---|---|
|  |  |  |  |
| 1 | 187 | 16,3 (18) | 6,2 |

| lepton masses | | | |
|---|---|---|---|
| 0,5 MeV/c² | 100 MeV/c² | 1800 MeV/c² | 14,4 GeV/c² |

**Fig. 11: Computational results for $p = 2,00$ after anharmonic correction.**

The result for the $m_{tau}/m_{mu}$ is now close enough to the experimental value to give an estimation for the mass of the next lepton to be found. To do so the equation set (9-22) has to be extended with an additional level ($(7/2)\tilde{h}\omega$), i.e. as:

$$k_0(d_0') + 3p_c d_0'^2 \Omega k_2(d_0') = k_0(d_a')$$

$$k_0(d_0') + 3p_c d_0'^2 \Omega k_2(d_0') = k_0(d_1') + p_{cc}\Omega d_1'^2 k_2(d_1')$$

$$k_0(d_1') + 5p_c d_1'^2 \Omega k_2(d_1') = k_0(d_b')$$

$$k_0(d_1') + 5p_c d_1'^2 \Omega k_2(d_1') = k_0(d_2') + p_{dd}\Omega d_2'^2 k_2(d_2')$$

$$k_0(d_2') + 7p_d d_2'^2 \Omega k_2(d_2') = k_0(d_c')$$

$$p_{dd} = 2p\left[\frac{3}{8} + \frac{1}{8\Omega^2} + \frac{3\beta}{\Omega^3}\right] \quad \text{and} \quad p_d = 2p\left[\frac{19}{8} + \frac{1}{8\Omega^2} + \frac{3\beta}{\Omega^3}\right]/5 \quad (9\text{-}42)$$

The results are shown in fig.11. This gives a prediction of about 12,2 GeV/c$^2$ for the next lepton to be found.

Note that the algorithmic fit requires a value of $p = 2,0$. Let us reconsider this parameter. It has been defined by (9-15). Knowing from experimental evidence that all vibrating mass $m'_p$ of a quark doublet eventually turns into bosonic mass $m'_W$, it is reasonable to expect that $m'_p$ can be equated with $m'_W$. As $m'_W$ is the boson mass observed from the center of the doublet and as the boson velocity is expected to be close to $c$, its value is high. This implies a high value for $m'_p$ as well, due to high-speed vibration. Therefore $m'_p$ should not be identified as a rest mass figure. However the simple harmonic oscillator model does not allow distinguishing between rest mass and relativistic mass. Therefore in our modeling increase of mass due to velocity has been accounted for by an effective mass ($m'_p$). So one might expect that the shortcomings of the simple modeling will show up in the factor $\alpha$ as defined in (9-7). Equating $m'_p$ with $m'_W$ in (9-15)

$$\alpha^2 = \frac{1}{p} = \frac{1}{2} \quad (9\text{-}43)$$

This result shows a good fit with the value $\alpha = \pi/4$ as predicted for the simple linear two-body oscillating system.

# 10. Conclusions

In the views as outlined in this paper we have taken the existence of the Higgs field as an undeniable starting point. We have however challenged the origin of the field. Rather than ascribing the origin of it to a yet undiscovered phantom particle, we have ascribed the origin directly to electromagnetic energy, in particular as magnetic charge next to electric charge of elementary poin-tlike particles. To this end we have used two instruments. The first one is the transform of the Higgs field from a functional description into a spatial description, without changing the basic properties. This is, as far as the author knows, not done before. The other instrument is as old as 1931. It is the concept of the magnetic monopole, as introduced by Dirac. The two instruments fit well together. In the paper it is shown that the particular field of the monopole as imposed by the Higgs field has prevented its experimental verification. As is well known from the work of Dirac, one of the consequences of the magnetic monopole is the discretization of electric charge. The result of all of this is that electromagnetic energy on its own is the source of all mass. It implies that the search after the Higgs particle will remain fruitless. No other equations, apart from Maxwell's Equations and Dirac's Equation are required to express the fundamentals of quantum waves and quantum fields, which makes the disputed Klein Gordon Equation obsolete. We have also shown that the theory as developed along these lines does not only make the neutrino less mysterious, but it also reveals an algorithm to explain the ratios between the lepton masses. In that sense the theory shows a predictive element, while grosso modo, as shown, no derogation is done to the results and instruments of canonic theory.

**Appendix A: Covariant derivative for local phase invariance**

From (3-6) and (3-7) we have:

$$\frac{D\Psi}{\partial x_i} = \exp[-j\vartheta(x)]\frac{D\Psi'}{\partial x_i} = \exp[-j\vartheta(x)]\left\{\frac{\partial}{\partial x_i}\Psi' + jqA_i'\Psi'\right\} \quad \text{(A-1)}$$

From (3-8) and (3-5):

$$\exp[-j\vartheta(x)]\frac{\partial}{\partial x_i}\Psi' = \frac{\partial\Psi}{\partial x_i} + j\Psi\frac{\partial}{\partial x_i}\vartheta(x) \quad \text{(A-2)}$$

Substitution of (A-2) into (A-1) gives:

$$\frac{D\Psi}{\partial x_i} = \frac{\partial\Psi}{\partial x_i} + j\Psi\frac{\partial}{\partial x_i}\vartheta(x) + jqA_i'\exp[-j\vartheta(x)]\Psi' \quad \text{(A-3)}$$

From (A-3) and (8-7) it follows that:

$$qA_i = qA_i' + \frac{\partial}{\partial x_i}\vartheta(x) \tag{A-4}$$

From (10-) and (10-) we have:

$$\frac{D\Psi}{\partial x_i} = \exp[-j\vartheta_k(x)]\frac{D\Psi''}{\partial x_i} = \exp[-j\vartheta_k(x)]\left\{\frac{\partial}{\partial x_i}\Psi'' + jgB_{ki}'\Psi''\right\} \tag{A-5}$$

From (10-) and (10-):

$$\exp[-j\vartheta_k(x)]\frac{\partial}{\partial x_i}\Psi'' = \frac{\partial\Psi}{\partial x_i} + j\Psi\frac{\partial}{\partial x_i}\vartheta_k(x) \tag{A-6}$$

Substitution of (A-6) into (A-5) gives:

$$\frac{D\Psi}{\partial x_i} = \frac{\partial\Psi}{\partial x_i} + j\Psi\frac{\partial}{\partial x_i}\vartheta_k(x) + jgB_{ki}'\exp[-j\vartheta_k(x)]\Psi'' \tag{A-7}$$

From (A-7) and (10-) it follows that:

$$gB_{ki} = gB_{ki}' + \frac{\partial}{\partial x_i}\vartheta_k(x) \tag{A-8}$$

**Appendix B: Expansion of the Ishii potential**

$$V(x) = f(x+d) + f(d-x) \text{ with } f(x) = \frac{\exp[-2x]}{x^2} - \frac{\exp[-x]}{x}$$

$$V(x) = g_2(x) - g_1(x)$$

$$g_2(x) = \frac{\exp[-2(x+d)]}{(x+d)^2} + \frac{\exp[-2(d-x)]}{(x-d)^2} = \frac{(d-x)^2\exp[-2(x+d)] + (x+d)^2\exp[-2(d-x)]}{(d^2-x^2)^2}$$

$$= \frac{\exp[-2d]}{(d^2-x^2)^2}\{(x^2-2dx+d^2)\exp[-2x]+(x^2+2dx+d^2)\exp[2x]\}$$

$$= \frac{\exp[-2d]}{(d^2-x^2)^2}\{(x^2+d^2)(\exp[-2x]+\exp[2x])+2dx(\exp[2x]-\exp[-2x])\}$$

$$\approx \frac{\exp[-2d]}{d^4}\left(1+\frac{x^2}{d^2}+\frac{x^4}{d^4}\right)\left(1+\frac{x^2}{d^2}+\frac{x^4}{d^4}\right)\left\{(x^2+d^2)\left(2+4x^2+\frac{4}{3}x^4\right)+8dx^2+\frac{16}{3}dx^4\right\}$$

$$\approx \frac{\exp[-2d]}{d^4}\left(1+\frac{2x^2}{d^2}+\frac{3x^4}{d^4}\right)\left\{2d^2+x^2(2+4d^2+8d)+x^4\left(4+\frac{4}{3}d^2+\frac{16}{3}d\right)\right\}$$

$$\approx \frac{\exp[-2d]}{d^4}\left\{2d^2+x^2(2+4d^2+8d)+4x^2+x^4\left(4+\frac{4}{3}d^2+\frac{16}{3}d\right)+\frac{2x^4}{d^2}(2+4d^2+8d)+6\frac{x^4}{d^2}\right\}$$

$$\approx \frac{\exp[-2d]}{d^4}\left\{2d^2+x^2(6+4d^2+8d)+x^4\left(12+\frac{10}{d^2}+\frac{16}{d}+\frac{4}{3}d^2+\frac{16}{3}d\right)\right\}$$

$$g_1(x) = \frac{(d-x)\exp[-(x+d)]+(x+d)\exp[-(d-x)]}{d^2-x^2} \approx$$

$$\approx \frac{\exp[-d]}{d^2-x^2}\{d(\exp[-x]+\exp[x])+x(\exp[x]-\exp[-x])\}$$

$$\approx \frac{\exp[-d]}{d^2-x^2}\left\{d\left(2+x^2+\frac{x^4}{12}\right)+2x^2+\frac{x^4}{3}\right\}$$

$$\approx \frac{\exp[-d]}{d^2}\left(1+\frac{x^2}{d^2}+\frac{x^4}{d^4}\right)\left\{d(2+x^2)+2x^2+x^4\left(\frac{d}{12}+\frac{1}{3}\right)\right\}$$

$$\approx \frac{\exp[-d]}{d^2}\left\{2d + x^2\left(2 + d + \frac{2}{d}\right) + x^4\left(\frac{2}{d^3} + \frac{2}{d^2} + \frac{1}{d} + \frac{1}{3} + \frac{d}{12}\right)\right\}$$

so: $$V(x) = g_2(x) - g_1(x) = k_0 + k_2 x^2 + k_4 x^4 + \ldots$$

with $$k_0 = 2\left(\frac{\exp[-2d]}{d^2} - \frac{\exp[-d]}{d}\right)$$

$$k_2 = \frac{\exp[-2d]}{d^4}(6 + 4d^2 + 8d) - \frac{\exp[-d]}{d^2}\left(2 + d + \frac{2}{d}\right)$$

$$k_4 = \frac{\exp[-2d]}{d^4}\left(\frac{10}{d^2} + \frac{16}{d} + 12 + \frac{16}{3}d + \frac{4}{3}d^2\right) - \frac{\exp[-d]}{d^2}\left(\frac{2}{d^3} + \frac{2}{d^2} + \frac{1}{d} + \frac{1}{3} + \frac{d}{12}\right)$$

## Side Note I: The relativistic wave equation neglecting spin.

The development of a spin-less limit for Dirac's Equation starts by observation of the equation set (2-18). Let $\Psi_1$ be dominant over $\Psi_2$. This implies a small value for the energy of $\Psi_2$. As we have from (2-1):

$$p_0^2 = p_x^2 - m_0 c^2$$

and as it is supposed that: $$m_0 c^2 \gg p_x^2$$

we have: $$p_0 = \pm j m_0 c$$

There are two options to choose, a minus sign and plus sign. If we opt for the minus sign, (2-18a) reduces to:

$$\hat{p}_x \Psi_1 \approx 2 j m_0 c \Psi_2. \tag{2-26}$$

After substitution of (2-26) into (2-18b) and subsequent evaluation we get after identifying $\Psi_1$ as $\Psi$.

$$j\tilde{h}\frac{\partial \Psi}{\partial t} + \frac{\tilde{h}^2}{2m_0}\frac{\partial^2 \Psi}{\partial x^2} - c^2\Psi = 0. \tag{2-27}$$

The non-relativistic limit of solutions of this equation appear to be solutions of Schrödinger's Equation:

$$j\tilde{h}\frac{\partial \Psi}{\partial t} + \frac{\tilde{h}^2}{2m_0}\frac{\partial^2 \Psi}{\partial x^2} = 0. \tag{2-28}$$

The chosen option for the minus sign corresponds with the negative temporal moment solution of Dirac's Equation. If we suppose that $\Psi_2$ is dominant over $\Psi_1$ we may follow the same procedure, but then with the positive temporal moment. The resulting wave equation is the same.

Evidently (2-27) is a valid relativistic version of Schrödinger's Equation. It does not show the flaws of the seriously disputed Klein-Gordon Equation [13,14,15,16] as, according to Dirac's requirement for positive definiteness, the temporal derivative is of first order, thereby obeying the condition for positive definiteness. The very reason for the difference with the Klein-Gordon Equation is the unjustified generalization of the basic hypothesis of quantum mechanics, as formulated in (2-3), of transforming the square of a momentum into a second order differential quotient on a wave function as done in the derivation of the Klein Gordon Equation.

## Side note II: the remnant field of a quark triplet (nucleon)

Let us now consider a triplet configuration of quarks as in baryons. We then have:

$$V_T[x, y, z] = f_1[x, y, z] + f_2[x, y, z] + f_3[x, y, z],$$

wherein

$$f_i[x, y, z] = \Phi_0 \frac{\exp[r_i]}{r_i}\left(\frac{\exp[r_i]}{r_i} - 1\right) \quad \text{with} \quad r_i = \lambda\sqrt{(x-x_i)^2 + (y-y_i)^2 + (z-z_i)^2}. \tag{1}$$

Let the three quarks be equally spaced at a distance $d$ apart from each other with the center of gravity in $(0, 0, 0)$. Calculation of the potential $V_T[0, 0, 0]$ in the symmetry point (0,0,0) results into:

$$V_T[0, 0, 0] = 3\left(\frac{e^{-2d'}}{d'^2} - \frac{e^{-d'}}{d'}\right). \tag{2}$$

This potential has a minimum $V_T[0, 0, 0] = -0,75$ for $d' = 0,852$ Fig.12 shows a 3D-plot and a contour plot of the potential in the plane of the quark centers. In fig.13 it is shown how the potential behaves along the axis $(0, 0, z)$ if the spacing parameter $d$ is changed.

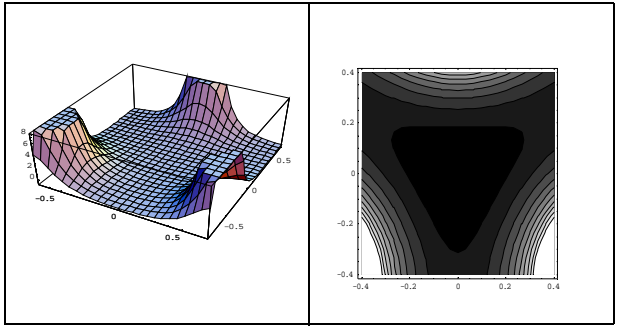

Fig. 12: 3D plot and contour plot of the potential of a quark triplet

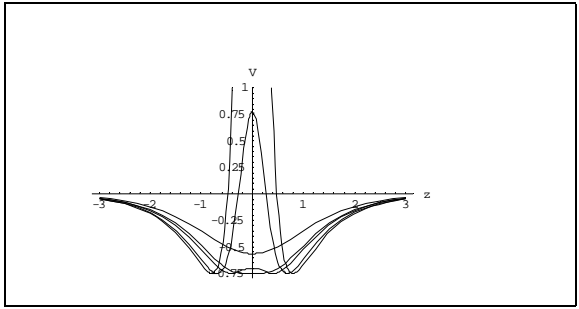

Fig. 13: Potential of the quark triplet along the axis (z,0,0). Shift to the right if the quark spacing decreases.

For $d' > 0,852$ the potential has a parabolic behavior above the minimum For $d' < 0,852$ up to a value of $d$ approaching zero the minimum is still there, but is now dual and shifted along the $z$-axis. Curiously the minimum value is not affected for whatever value of $d' < 0,852$. This property resembles the asymptotic freedom in the canonic quark theory. Let us address now the question how two of those triplets will behave if they are brought into close encounter. Let us do so be considering the behavior of the potential function along the axis $(0,0,z)$. From (2) it follows that this potential $V_z(z)$ is given by:

$$V_z(z) = 3 \frac{\exp[-\lambda\sqrt{d^2+z^2}]}{\lambda\sqrt{d^2+z^2}} \left( \frac{\exp[-\lambda\sqrt{d^2+z^2}]}{\lambda\sqrt{d^2+z^2}} - 1 \right). \tag{3}$$

Let us further consider two triplets in opposite planar orientation at a vertical distance 2$a$ apart. The potential function $V_{2z}(z)$ of this doublet can thus be written as:

$$V_{2z}(z) = V_z(a+z) + V_z(a-z). \tag{4}$$

Elementary algebraic analysis of (4) under the condition of $a \gg d$ shows compatibility of (3) and (4) with:

$$V_z(z) = 3\Phi_0 \frac{\exp[-\lambda z]}{\lambda z}\left(\frac{\exp[-\lambda z]}{\lambda z} - 1\right). \tag{5}$$

So, if the quarks within the nucleon are closely spaced, in the way as we have related above with asymptotic freedom, the composite potential function along the *z*-axis at value $z \gg d$ of three quarks shows the same behavior as that of a single quark  Once understood, it will not surprise that the restriction "along the *z*-axis" can be dropped. At sufficient distance the triplet's behavior is omni-directionally similar to that of a single quark. It is a straightforward explanation for the Van der Waals-type remnant field.

Similarly as discussed in the previous section the potential fields of two nucleons spaced at short distance will click together into a doublet field with field characteristics of a quantum mechanical oscillator. The nucleons are bound by force carrying particles. Clearly in this case these bosons are the well known pi-mesons, which are emitted or absorbed under change of energetic state of the nucleon doublet.  Usually the boson mechanism is related with the decay mechanism of particles rather than with a mechanism of quantum mechanical oscillation. Decay mechanisms are an exponential function of time. Let $N(t)$ the total number of particles radiated from a centric source. The number $N(t)$ of particles in a wave front after time *t* is:

$$N(t) = N_0 \exp[-\gamma t] \tag{6}$$

wherein $\gamma$ is the decay rate. If *v* is the velocity of the particles then the number remaining at distance *r* is:

$$N(t) = N_0 \exp[-(\gamma/v)r]. \tag{7}$$

So, the force executed by bosons (force carrying particles) per unit of area is:

$$F(r) \sim \frac{\exp[-(\gamma/v)r]}{r^2} \tag{8}$$

This exponential force behavior corresponds with the  force behavior as a consequence of the inter nucleon potential as expressed by (5) in the region of the attractive force. As

$$V(r) \approx \Phi_0\left(\frac{\exp[-\lambda r]}{\lambda r}\right) \tag{9}$$

this force behavior is:

$$F(r) \sim \frac{\exp[-\lambda r]}{r^2}. \tag{10}$$

This gives the following relationship

$$\lambda = \gamma/v. \tag{11}$$

Let us compare the role of bosons in the decay mechanism with the role of bosons as carriers of force in an equilibrium of a doublet of particles, be it a nucleon doublet or a quark-antiquark doublet. In this latter case particles hold each other in a vibrating balance of ground state energy. If the energy of vibration increases the virtual particles are emitted and become observable. The decay mechanism at the other hand allows to consider the bosons as virtual radiation from a separate (field creating) particle (singlet). Both views are complementary and don't exclude each other.

So far it has been tacitly supposed that the primary parameters $\lambda$ and $\Phi_0$ are invariant properties of a quark. If it is true, observations as for instance relationships with life time, or other ones, should invariantly hold for all boson types. Let us compare the nucleon spacing in the nucleon doublet with the quark spacing in the pi-meson. Assuming as before that

$$m'_\pi = E_\pi = \alpha_0 \frac{\tilde{h} v_\pi}{d_\pi} \quad \text{and similarly} \quad m'_W = E_W = \alpha_0 \frac{\tilde{h} v_W}{d_W} \tag{12}$$

(with $\alpha_0 = \pi/2$) we have:

$$\frac{m'_W}{m'_\pi} = \frac{d_\pi v_W}{d_W v_\pi}. \tag{13}$$

Herein are $m'_W$ and $m'_\pi$ the energetic figures for the masses of respectively the W-boson and the pi-meson as observed from the center of the quark doublet. As the observer in this center sees both masses moving with respective velocities $v_W$ and $v_\pi$, we may rewrite this expression in terms of rest masses as:

$$\frac{m'_{0W}}{m'_{0\pi}} \sqrt{\frac{1-(v_\pi/c)^2}{1-(v_W/c)^2}} = \frac{v_W}{v_\pi}. \tag{14}$$

Expressing the velocities as fraction of the light velocity, i.e. by: $\alpha = v_W/c$ and $\beta = v_\pi/c$ (14) can be rewritten as:

$$\frac{m'_{0W}}{m'_{0\pi}} \sqrt{\frac{1-\beta^2}{1-\alpha^2}} = \frac{\alpha}{\beta}. \tag{15}$$

Evaluation of (15) results into the following condition:

$$\alpha^2 = \frac{1-\sqrt{1-4Q}}{2} \quad \text{with} \quad Q = \left(\frac{m'_{0W}}{m'_{0\pi}}\right)^2 \beta^2 (1-\beta^2). \tag{16}$$

As $m'_{0W}/m'_{0\pi} \gg 1$ real solutions for $\alpha$ are only possible if $\beta$ is close to 1. So at least the pi-meson velocity has to be close to the light velocity. Is it possible that $\alpha$ is close to the light velocity as well? To investigate so we define:

$$1-\beta^2 = \Delta_\pi \quad \text{and} \quad 1-\alpha^2 = \Delta_W. \tag{17}$$

Insertion of (17) into (15) and subsequent evaluation yields:

$$\frac{\Delta_\pi}{\Delta_W} = \frac{\alpha^2}{\beta^2}\left(\frac{m'_{0W}}{m'_{0\pi}}\right)^2 = \frac{(1-\Delta_\pi)}{(1-\Delta_W)}\left(\frac{m'_{0\pi}}{m'_{0W}}\right)^2, \tag{18}$$

so that under the condition $\Delta_W \ll 1$, considering that $\Delta_\pi \ll 1$ we have for (18):

$$\frac{\Delta_\pi}{\Delta_W} \approx \left(\frac{m'_{0\pi}}{m'_{0W}}\right)^2. \tag{19}$$

Therefore both the pi-meson velocity and the W-boson velocity can be close to the light velocity as long as the pi-meson velocity is much closer to it than the W-boson velocity. Let us now compare the nucleon spacing in the nucleon doublet with the quark spacing in the pi-meson. Note that expressions (12,13) hold for an observer in the center of the doublets. For the lab frame observer we have:

$$\left(\frac{d_W}{d_\pi}\right)_{lab} = \frac{v_W m'_{0\pi}}{v_\pi m'_{0W}} \sqrt{\frac{1-(v_W/c)^2}{1-(v_\pi/c)^2}} \approx \frac{m'_{0\pi}}{m'_{0W}} \sqrt{\frac{\Delta_W}{\Delta_\pi}}, \tag{20}$$

so that with (19): $\left(\frac{d_W}{d_\pi}\right)_{lab} \approx 1. \tag{21}$

So the conclusion is that the lab frame observer experiences near light velocity for pi-mesons and W-bosons and about the same spacing between the constituents of both doublets.

# Supplement: On the relationship between Gravity and Electromagnetism

**Summary**
In this paper a quantum mechanical equivalent is derived for Einstein's Field Equation. This is done by calculating the curvature of space time under influence of the Higgs field and relating the result with the energy momentum content in it. The proportionality constant between the two is the equivalent of the gravitational constant. It is discussed whether this equivalent is equal to the gravitational constant itself in an attempt to unify quantum mechanics, electromagnetism and gravity. The numerical value of the derived quantum mechanical formula for the gravitational constant appears to fit with its well known value.

## 1.0 Introduction

Previously we have described the quark-antiquark doublet as a cloud of vibrating (electro)magnetic energy [1,2]. This vibration has been modeled as the motion of a particle in a energetic field with potential characteristics similar to an (an)harmonic quantum mechanical oscillator. This induces the idea that somehow it must be possible to relate relativistic massive motions with electromagnetism. Motions under general relativity are captured by the concept of Gravity. The relationship as mentioned would be manifest if it would be possible to *calculate* the gravitational constant $G$ out of parameters belonging to the Quantum Theory. Would that be possible? This is the issue we wish to address in this document.

We shall do so along the following lines. In the next chapter (chapter 2) we first summarize the leading considerations to outline the approach. This will show the need for deriving a quantum mechanical wave function of a particle in terms of the metric tensor of the energetic space time in which it moves. This will be done in chapter 3. In chapter 4 the metric tensor will be expressed in terms of quantum mechanical parameters, in particular in the parameters of the so-called Higgs field. The reason to do so is, that, as shown in the present state of Quantum Theory, the Higgs field is the link between electromagnetism and quantum mechanics. In chapter 5 the ingredients of Einstein's Field Equation (which in fact captures the Gravity Theory), i.e. the Einstein tensor and the energy momentum tensor, are formulated in respectively parameters of the Higgs field (replacing the metric tensor parameters) and energetic parameters of a quantum wave. The new formulation of Einstein's Field Equation will yield an expression for the equivalent of the gravitational constant in terms of electromagnetic and quantum wave parameters.

Is this approach fundamentally different from the wrong and naive approach to equate Coulomb's law with Newton's gravity law? In that case the physical mass is calculated of two charged particles out of the energies of their electric fields and the attractive Newtonian force between the resulting masses is equated with the attractive or repulsive electric force in order to obtain an expression for the gravitational constant. The outcome of this is false by an order of magnitude of about $4,3 \times 10^{38}$. So it is fair to ask whether the approach described above is not a complex way for the same approach and therefore an intellectual trap once more. However where the naive approach equates a strong bipolar (electric) force with a weak unipolar (gravitational) force the method described above equates two unipolar forces. This is the basic reason why the two approaches are fundamentally different indeed.

## 2.0 Considerations and motivation

Let us start with summarizing the considerations outlining the approach for analysis:

1. The motion of a massive test particle in a field of forces is subject to the Principle of Equivalence. This principle says that the particle's motion under force can be derived from its free space motion by a simple transform rule. The transform rule is a transform of coordinates of space time. The resulting equation of motion is the Geodesic Equation, which contains the metric tensor ($g_{ij}$) as a parameter. The metric tensor describes the characteristics of the transform of coordinates.

2. The quantum mechanical wave equation of a massive particle in free space can be derived by its equation of motion in free space by a simple transform rule. The rule consists of transforming momenta into first order differentiation of a wave function.

3. The quantum mechanical wave equation of a massive particle in a field of forces is subject to Yang Mills' Principle. This principle says that the particle's wave function under force can be derived from its free space wave function by a simple transform rule. The rule is a change of global phase invariance into local phase invariance, similarly as the change of global Lorentz transform into local Lorentz transform in relativistic motions.

4. Electromagnetic fields are formats of energy, similarly as massive particles in rest. All formats of energy are subject to Einstein's Field Equation. This equation is expressed in parameters that can be derived from the metric tensor at one end: the Einstein tensor part ($G_{ij}$). At the other end the equation contains parameters that can be derived from energy as present in the space as characterized by the metric tensor: the energy momentum part ($T_{ij}$). There is a proportionality factor $G$ involved, known as the gravity constant. The full expression is:

$$G_{ij} = \frac{8\pi G}{c^4} T_{ij} \quad \text{with} \quad G_{ij} = R_{ij} - \frac{1}{2} R g_{ij}. \tag{2-1}$$

Herein $R_{ij}$ and $R$ are respectively the so-called Ricci tensor and the Ricci scalar which can be calculated if the metric tensor components $g_{ij}$ are known.

5. The equivalent of the energy momentum tensor of a cloud of electromagnetic energy ($T_{ij}^{EM}$) can be expressed in terms of electric fields $E$ and magnetic fields $H$ [3]:

$$\text{if } i = 0 \text{ and } j = 0: \quad T_{00}^{EM} = \frac{1}{2}(\varepsilon_0 E^2 + \mu_0 H^2)$$

$$\text{if } i = 0 \text{ and } j \neq 0: \quad T_{0j}^{EM} = (\mathbf{E}\sqrt{\varepsilon_0} \times \mathbf{H}\sqrt{\mu_0})_j$$

$$\text{if } i \neq 0 \text{ and } j \neq 0: \quad T_{ij}^{EM} = \varepsilon_0 E_i E_j + \mu_0 H_i H_j - \frac{1}{2}\delta_{ij}(\varepsilon_0 E^2 + \mu_0 H^2), \text{ wherein}$$

$$E^2 = \sum_{j=1}^{3} E_j^2, \quad H^2 = \sum_{i=1}^{3} H_j^2 \text{ and } \delta_{ij} = 1 \text{ if } i = j \text{ and } \delta_{ij} = 0 \text{ if } i \neq j. \tag{2-2}$$

What kind of conclusions can be drawn from these considerations? Let us suppose that the dynamics of an electromagnetic cloud can be described as a test particle moving in a space described by a metric tensor. Its quantum mechanical wave equation can be obtained by transform of the equation of motion (consideration 2). This wave equation will contain the parameters of the metric tensor. That is one way to obtain the wave equation. But there is another way as well. This other way is to extract the wave equation from Yang Mills principle (consideration 3). This wave function will contain parameters of a potential field. Knowledge of this potential means that the metric tensor parameters can be extracted in terms of this potential field. Equating the two expressions of the quantum mechanical wave equation can do this. This will enable to calculate the Einstein tensor ($G_{ij}$) in terms of parameters of the potential field. See (2-1). The potential field however is an expression of an electromagnetic behavior. It means that we have expressed now the left hand part of Einstein's Field Equation ($G_{ij}$) in terms of parameters of the electromagnetic field.

What about the right hand part? Apart from the proportionality factor involving the gravitational constant $G$, the right hand part consists of the energy momentum tensor, which we can express in terms of the electromagnetic field as well (consideration 5). As soon as these two parts of the Einstein Field Equation can be retrieved, the proportionality $G$ can, at least in principle, be calculated. So, the crux is in the knowledge of the potential field.

How to obtain this knowledge? In a previous paper we have shown that a quark doublet can be described as a one-dimensional cloud of vibrating (electro)magnetic energy [2]. This energy has been modeled as a test particle moving in an energetic field with potential characteristics similar to an harmonic or anharmonic quantum mechanical oscillator. Moreover we have shown that this energetic field can be identified as the Higgs field described in terms of spatial parameters [2]. All these ingredients indicate the possibility to relate the gravitational constant via the Higgs field with electromagnetism. That would mean unification between gravity, quantum mechanics and electromagnetism.

Needless to say that this would be a striking conclusion, because the unification problem is the most outstanding problem of to-day's physics.

The question is merely how the mathematics to do so can be organized. This will be the subject of the subsequence of this paper. First of all we wish to derive a quantum mechanical wave equation in terms of components of the metric tensor (section 3). After that we wish to express the metric tensor in terms of quantum mechanical parameters (section 4). Equating two versions of the quantum mechanical wave equation: one including the metric tensor and another one including the Higgs field will do this. In section 5 the electromagnetic energy momentum tensor of the Higgs field will be derived. Using these results, in section 6 an equivalent of the gravitational constant will be expressed in terms of electromagnetic parameters.

## 3.0 The wave function in terms of the metric tensor.

In this section we wish to obtain a quantum mechanical wave equation in terms of the components of the metric tensor. This can be done by transform of the equation of motion of a test particle in a field of forces. This equation of motion is known as the geodesic equation. Fortunately the space time under test has a single spatial dimension. This is a result of the modeling of the quark-antiquark doublet by a linear mechanical motion. Under this condition the equation of motion can be read as [4][1]:

$$\frac{d^2 x}{d\tau^2} + \frac{1}{2g_{xx}}\left[\frac{\partial g_{xx}}{\partial x}\left(\frac{dx}{d\tau}\right)^2 - \frac{\partial g_{tt}}{\partial x}\left(\frac{dt'}{d\tau}\right)^2 + 2\frac{\partial g_{xx}}{\partial t'}\left(\frac{dx\,dt'}{d\tau\,d\tau}\right)\right] = 0$$

$$\frac{d^2 t'}{d\tau^2} + \frac{1}{2g_{tt}}\left[\frac{\partial g_{tt}}{\partial t'}\left(\frac{dt'}{d\tau}\right)^2 - \frac{\partial g_{xx}}{\partial t'}\left(\frac{dx}{d\tau}\right)^2 + 2\frac{\partial g_{tt}}{\partial x}\left(\frac{dx\,dt'}{d\tau\,d\tau}\right)\right] = 0 \quad . \tag{3-1a,b}$$

What does it mean? We see the following parameters: $x$, $\tau$, $t'$, $g_{xx}$ and $g_{tt}$. Of course $x$ is the spatial coordinate. Furthermore we have two different time coordinates. The parameter $t'$ is the normalized time for the stationary observer, i.e. $t' = jct$, wherein $c$ is the light velocity and $j = \sqrt{-1}$. The parameter $\tau$ is proper time, i.e. the wrist time of a co-moving observer (co-moving with the particle). The parameters $g_{xx}$ and $g_{tt}$ are elements of the so-called metric tensor. They determine the way how the frame of coördinates $\xi$, $\tau'$ of the co-moving observer is transformed (by considering $\xi(x, t')$ and $\tau'(x, t')$ into the frame of coordinates $(x, t')$ of the stationary observer. In particular:

---

1. This equation is the "1+1-dimensional" form of Einstein's Geodesic Equation. It requires some effort to simplify the Geodesic Equation for a single spatial dimension. It is simpler to set-up the equation directly in a single spatial dimension and to check the result against Einstein's full format. Unfortunately the author does not know any reference apart from [4] where it is done so.

$$\left(\frac{\partial \xi}{\partial x}\right)^2 + \left(\frac{\partial \tau'}{\partial x}\right)^2 = g_{xx} \quad \text{and} \quad \left(\frac{\partial \xi}{\partial t'}\right)^2 + \left(\frac{\partial \tau'}{\partial t'}\right)^2 = g_{tt}, \text{ with } \tau' = jc\tau. \tag{3-2a,b}$$

The geodesic equation is a result of Einstein's Principle of Equivalence that states that the forced movement of the particle seen by the stationary observer is seen as a free movement seen by the co-moving observer. Therefore the geodesic equation is nothing more than a transformation of the free movement equation by merely transforming the coordinates.

Note that the time parameter of the geodesic equation is the proper time. Note also that temporal and spatial parameters occur on par, i.e. both equations are fully symmetrical. This is a basic feature of relativistically invariant equations.

The parity is lost after introduction a basic step in further evaluation. This step is the assumption of stationary. This means that we shall assume that the elements of the metric tensor are independent of time. This has to do with the nature of the energy field in which the particle is supposed to move. In many cases, such as motions in free space, motions in a gravity field or in a static electromagnetic field, this assumption is justified. If however this static character is violated, for instance as a consequence of particular forms of particle interaction, a reconsideration will be necessary. Such reconsideration is beyond the scope of this paper.

Because of the stationary condition the geodesic equation simplifies to:

$$\frac{d^2 x}{d\tau^2} + \frac{1}{2g_{xx}}\left[\frac{dg_{xx}}{dx}\left(\frac{dx}{d\tau}\right)^2 - \frac{dg_{tt}}{dx}\left(\frac{dt'}{d\tau}\right)^2\right] = 0$$

and
$$\frac{d^2 t'}{d\tau^2} + \frac{1}{g_{tt}}\frac{dg_{tt}}{dx}\left(\frac{dx}{d\tau}\frac{dt'}{d\tau}\right) = 0. \tag{3-3a,b}$$

The latter equation can be easily integrated, resulting in a linear differential equation:

$$\frac{dt'}{d\tau} = \frac{k_{int}}{g_{tt}}, \text{ wherein } k_{int} \text{ is an integration constant to be determined.} \tag{3-4}$$

In addition we may consider a redundant equation that makes further use of (3-3a) obsolete. This equation is the formulation of the invariance of the local space time interval, i.e.:

$$d\tau'^2 = g_{xx}dx^2 + g_{tt}dt'^2 \text{ so: } \left(\frac{d\tau'}{d\tau'}\right)^2 = g_{xx}\left(\frac{dx}{d\tau'}\right)^2 + g_{tt}\left(\frac{dt'}{d\tau'}\right)^2, \tag{3-5}$$

which is equivalent with
$$g_{xx}\left(\frac{dx}{d\tau}\right)^2 + g_{tt}\left(\frac{dt'}{d\tau}\right)^2 = -c^2. \tag{3-6}$$

Applying (3-6) to (3-4) we get:

$$\left(\frac{dx}{d\tau}\right)^2 = \frac{1}{g_{xx}}\left[-c^2 - \frac{k_{int}^2}{g_{tt}}\right]. \tag{3-7}$$

It can be shown that elaboration on the basis of (3-3a) instead of (3-6) gives the same result. Equations (3-4) and (3-7) together form an excellent set to derive wave equations for a single spatial dimension by basic rules in which momenta are transformed into operators on wave functions.

If we formulate (3-7) and (3-4) in momentum notation we have:

$$\frac{p_0}{m_0} = \frac{k_0}{g_{tt}} \quad \text{and} \quad \left(\frac{p_x}{m_0}\right)^2 = \frac{1}{g_{xx}}\left[-c^2 - \frac{k_{int}^2}{g_{tt}}\right], \text{ wherein } \frac{p_0}{m_0} = \frac{dt'}{d\tau} \text{ and } \frac{p_x}{m_0} = \frac{dx}{d\tau}. \tag{3-8a,b}$$

From (3-8b) it follows that:

$$\frac{p_x}{m_0} = \pm\sqrt{\frac{1}{g_{xx}}\left[-c^2 - \frac{k_{int}^2}{g_{tt}}\right]}. \tag{3-9}$$

We now invoke the basic quantum mechanical hypothesis describing transform of momenta into operations on a wave function such that:

$$p_i \to \hat{p}_i \Psi \quad \text{with} \quad \hat{p}_i = \frac{\tilde{h}}{j}\frac{\partial}{\partial x_i}. \tag{3-A}$$

From (3-A) and (3-8) we obtain operator and differential equations:

$$\hat{p}_0 \Psi = \frac{\tilde{h}}{j}\frac{\partial \Psi}{\partial x_0} = \frac{\tilde{h}}{j}\frac{\partial \Psi}{\partial jct} = -\frac{\tilde{h}}{c}\frac{\partial \Psi}{\partial t} = \frac{m_0 k_{int}}{g_{tt}}\Psi \quad \text{so:} \quad j\tilde{h}\frac{\partial \Psi}{\partial t} = -jcm_0\frac{k_{int}}{g_{tt}}\Psi$$

and

$$\hat{p}_x \Psi = \frac{\tilde{h}}{j}\frac{\partial \Psi}{\partial x} = \frac{\tilde{h}}{j}\frac{\partial \Psi}{\partial x} = \frac{\tilde{h}}{j}\frac{\partial \Psi}{\partial x} = \pm\sqrt{\frac{1}{g_{xx}}\left[-c^2 - \frac{k_{int}^2}{g_{tt}}\right]}\Psi \quad \text{so:}$$

$$\frac{\tilde{h}^2}{m_0}\frac{\partial \Psi}{\partial x} = \pm j\tilde{h}\sqrt{\frac{1}{g_{xx}}\left[-c^2 - \frac{k_{int}^2}{g_{tt}}\right]}\Psi. \tag{3-10}$$

As the curvature is assumed to be stationary we have:

$$j\tilde{h}\frac{\partial \Psi}{\partial t} = f_1(x)\Psi \quad \text{and} \quad \frac{\tilde{h}^2}{m_0}\frac{\partial \Psi}{\partial x} = \pm jf_2(x)\Psi$$

$$\text{with } f_1(x) = -jcm_0\frac{k_{int}}{g_{tt}} \text{ and } f_2(x) = \tilde{h}\sqrt{\frac{1}{g_{xx}}\left[-c^2 - \frac{k_{int}^2}{g_{tt}}\right]}. \quad (3\text{-}11\text{a,b})$$

It is our aim to relate the metric tensor components $g_{tt}$ and $g_{xx}$ with a potential field. As potential field is not a valid concept in General Relativity this is not trivial. It requires establishing a relationship of the relativistic wave equation set as expressed by (3-11) with the non-relativistic Schrodinger equation. This is possible under the adoption of some simplifications. Apart from stationarity (implying that both $g_{tt}$ and $g_{xx}$ are independent of time), we shall adopt two other conditions. These are:

(a) *isotropy and weak field* implying that:

$$g_{tt}(x) = g_{xx}(x) = g_{00}(x) \quad \text{and} \quad g_{00} = 1 + \Delta(x) \quad \text{with} \quad |\Delta(x)| \ll 1, \quad (3\text{-}12)$$

and (b) *non relativistic limit*, implying that

$$v_x/c \ll 1. \quad (3\text{-}13)$$

If (3-11b) is differentiated with respect to $x$ we get:

$$\frac{\partial^2 \Psi}{\partial x^2} = \frac{-m_0^2 f_2^2(x)}{\tilde{h}^2}\Psi \pm jm_0\frac{\partial f_2(x)}{\partial x}\Psi. \quad (3\text{-}14)$$

Under the conditions as mentioned we get:

$$f_1(x) \approx m_0c^2(1-\Delta) \quad f_2(x) \approx j\frac{c}{\tilde{h}}\sqrt{\Delta} \quad \text{and} \quad \frac{df_2}{dx} \approx j\frac{c}{2\tilde{h}\sqrt{\Delta}}\frac{d\Delta}{dx}. \quad (3\text{-}15)$$

As $|\Delta(x)| \ll 1$ and $v_x/c \ll 1$ we have from (3-4):

$$\frac{dt'}{d\tau} \approx jc \quad (3\text{-}16)$$

so that from (3-A) and (3-11a):

$$j\tilde{h}\frac{\partial \Psi}{\partial t} \approx m_0 c^2 \Psi. \tag{3-17}$$

Furthermore we may state in (3-14) that:

$$\left|\frac{-m_0^2 f_2^2(x)}{\tilde{h}^2}\right| \ll \left|m_0 \frac{\partial f_2(x)}{\partial x}\right| \tag{3-18}$$

so for (3-14) it may be written that:

$$\frac{\partial^2 \Psi}{\partial x^2} \approx \pm j m_0 \frac{\partial f_2(x)}{\partial x} \Psi. \tag{3-19}$$

Equations (3-17) and (3-19) together compose a non-relativistic wave equation. It enables solutions of the format $\Psi = \Psi_t \Psi_x$. The time independent part as obtainable from (3-19) can be related with the time independent part of Schrodinger's equation as it can now be stated that

$$\frac{\tilde{h}^2}{2m_0}\frac{d^2 \Psi_x}{dx^2} = V_r(x)\Psi \text{ with } V_r(x) = \pm\frac{\tilde{h}c}{2}\frac{dy}{dx} \text{ and } y = \sqrt{\Delta} \tag{3-20a,b}$$

This result means that a relationship has been established between a potential field $V_r(x)$ and the components of the metric tensor. If $V_r(x)$ is known, $\Delta$ can be calculated from (3-20b), thereby establishing $g_{00}$ and $g_{xx}$ via (3-12).

## 4.0 The metric tensor in terms of quantum mechanical parameters.

In the previous section we have obtained a quantum mechanical wave equation in terms of the metric tensor, albeit that the metric tensor is indirectly expressed in a parameter $y$. This parameter $y$ can be easily translated into $g_{ij}$ as we have from (3-17), (3-18) and (3-21):

$$g_{tt}(x) = g_{xx}(x) = 1 + y^2. \tag{4-1}$$

How to relate this with electromagnetism? We do so by invoking the model of the quark doublet [1]. This model describes the motion of a particle with effectiver mass $m_p$ in a Higgs field. This field is supposed to be of magnetic origin and the particle is supposed to be sensitive only for the magnetic force. Two parameters, a first one $\Phi_0$ with energetic dimension and a second one $\lambda$ with a dimension m$^{-1}$, characterize the field. The energetic parameter can be expressed in terms of magnetic field energy. The time independent part of the wave equation of the particle is:

$$\frac{\tilde{h}^2}{2m_{eff}}\frac{\partial^2 \Psi}{\partial x^2} = V_H(x)\Psi \quad \text{with } V_H(x) = \Phi_0\{k_0 + k_2(x\lambda)^2 + ...\}. \quad (4\text{-}1)$$

Equations (3-20a) and (4-1) are simultaneously true if $m_0 = m_{eff}$ and if

$$\pm\frac{\tilde{h}c}{2}\frac{dy}{dx} = \Phi_0\{k_0 + k_2(x\lambda)^2 + ...\}. \quad (4\text{-}2)$$

Integration of this equation yields:

$$y = \pm\frac{2\Phi_0}{\tilde{h}c}\left(k_0 x + \frac{2}{3}k_2\lambda^2 x^3 + ...\right). \quad (4\text{-}3)$$

Under consideration of (3-31):

$$\Delta \approx \left(\frac{2\Phi_0}{\tilde{h}c}\right)^2\left(k_0^2 x^2 + \frac{4}{3}k_0 k_2 \lambda^2 x^4 + ...\right). \quad (4\text{-}4)$$

so, under consideration of (3-18):

$$g_{00} = g_{11} \approx 1 + \left(\frac{2\Phi_0}{\tilde{h}c}\right)^2\left(k_0^2 x^2 + \frac{4}{3}k_0 k_2 \lambda^2 x^4 + ...\right). \quad (4\text{-}5)$$

That means that we have expressed now the metric tensor in terms of the Higgs field. The Higgs field is characterized by two parameters, i.e. $\lambda$ and $\Phi_0$. Note that $k_0$ (= -1/2) and $k_2$ (= 2,62) are no parameters, but known constants. Note also that the Higgs field parameters are parameters of the spatial representation of the Higgs field. Canonically the Higgs field is expressed as a function of the wave function, which shows two parameters as well. The spatial expression however allows a more easy interpretation of the Higgs field [1]. In a subsequent section we wish to assign quantitative values to the parameters ($\lambda$) and $\Phi_0$.

## 5.0 The Einstein tensor and the energy momentum tensor of the Higgs field.

### 5.1 The Einstein tensor

The results of the preceding section enables us to express the Einstein tensor ($G_{ij}$) of Einstein's Field Equation (2-1) in terms of the Higgs field parameters $\lambda$ and $\Phi_0$. The Einstein tensor is defined as:

$$G_{ij} = R_{ij} - \frac{1}{2} R g_{ij}. \tag{5-1}$$

In case of Cartesian space time with a single spatial dimension rather simple expressions for the Ricci tensor $R_{ij}$ and the Ricci scalar $R$ are obtained. For the Ricci tensor we may write [5,4][1]

$$R_{tt} = -\frac{1}{2}\frac{g_{tt}''}{g_{xx}} + \frac{1}{4}\frac{g_{tt}'}{g_{xx}}\left(\frac{g_{tt}'}{g_{xx}} + \frac{g_{tt}'}{g_{tt}}\right), \quad R_{xx} = -\frac{1}{2}\frac{g_{tt}''}{g_{tt}} + \frac{1}{4}\frac{g_{tt}'}{g_{tt}}\left(\frac{g_{xx}'}{g_{xx}} + \frac{g_{tt}'}{g_{tt}}\right), \quad R_{tx} = R_{xt} = 0. \tag{5-2}$$

(Note: the symbols ' and " stand for differentiation and double differentiation after the spatial parameter *x*)

Under the symmetry condition (to be discussed below) we have:

and therefore:
$$R_{tt} = R_{xx} = R_{00} = -\frac{1}{2}\frac{g_{00}''}{g_{00}} + \frac{1}{2}\frac{g_{00}'^2}{g_{00}^2}. \tag{5-3}$$

The Ricci scalar is:

$$R = \sum_{i=0}^{1}\sum_{j=0}^{1} g^{ij} R_{ij} = 2 g_{00} R_{00}. \tag{5-4}$$

Note: $g^{ij}$ is the inverse matrix of $g_{ij}$. From (5-3) and (4-5) it follows that:

$$g_{00}' \approx 2 k_0^2 x \left(\frac{2\Phi_0}{\tilde{h}c}\right)^2 \qquad g_{00}'' \approx 2 k_0^2 \left(\frac{2\Phi_0}{\tilde{h}c}\right)^2 \tag{5-5}$$

so:

$$R_{00} \approx -k_0^2 \left(\frac{2\Phi_0}{\tilde{h}c}\right)^2 + \frac{1}{2} k_0^4 x^2 \left(\frac{2\Phi_0}{\tilde{h}c}\right)^4. \tag{5-6}$$

For the Einstein tensor we may write under consideration of (5-4) and (3-18):

---

1. This expression, or equivalent, can be found in many textbooks on General Relativity. The format as shown here is most easily recognized in [5,4]

$$G_{00} = R_{00} - \frac{1}{2}Rg_{00} = R_{00}(1 - g_{00}^2) = R_{00}\{1 - (1 + 2\Delta)\} = -2\Delta R_{00}, \qquad (5\text{-}7)$$

wherein $\Delta$ is given by (4-4). From (5-6) and (5-7) we obtain:

$$G_{00} \approx 2k_0^4 \left(\frac{2\Phi_0}{\tilde{h}c}\right)^4 x^2. \qquad (5\text{-}8)$$

## 5.2. The energy momentum tensor

The right-hand part of Einstein's Field equation is described by the energy momentum tensor. The electromagnetic equivalent determines electromagnetic field density. In the Higgs field the magnetic field dominates largely over the electric field (if there is any). The calculation from (2-2) shows that:

$$T_{00} = T_{11} = \frac{1}{2}\mu_0 H^2(x) \quad \text{and} \quad T_{10} = T_{01} = 0. \qquad (5\text{-}9)$$

Inspection of Einstein's Field Equation under consideration of (5-9) and (5-2) shows that the equation can be satisfied for all $x$ if and only if

$$g_{tt}(x) = g_{xx}(x) = g_{00}(x). \qquad (5\text{-}10)$$

That means that the previously adopted symmetry condition is fulfilled indeed.

From [1] we know that:

$$B(x) = \frac{g}{cq_e}\frac{d\Phi}{dx} \quad \text{wherein} \quad q_e^2 = 4\pi\varepsilon_0 \tilde{h}cg^2. \qquad (5\text{-}11\text{a,b})$$

and $B = \mu_0 H$, $g$ the universal quantum mechanical coupling and $q_e$ the elementary electric charge. As $\Phi(x)$ is the Higgs field as expressed by (4-1), we have

$$B(x) = \frac{g\Phi_0}{cq_e}(2k_2\lambda^2 x + 4k_4\lambda^4 x^3 + \ldots) = \mu_0 H(x) \qquad (5\text{-}12)$$

so with (5-9):

$$T_{00} = \frac{1}{2}\mu_0 H^2(x) = \frac{1}{2}\mu_0\left(\frac{g\Phi_0}{\mu_0 cq_e}\right)^2 (4k_2^2\lambda^4 x^2 + \ldots) \qquad (5\text{-}13)$$

Under consideration of (5-11b) we have

$$T_{00} \approx k_2^2 \lambda^4 x^2 \frac{\Phi_0^2}{2\pi \tilde{h} c} .\tag{5-14}$$

## 6.0 The gravitational constant

It is now possible to express the gravitational constant in terms of the quantum mechanical parameters $\lambda$ and $\Phi_0$. As is obvious from (5-12) these parameters are in fact electromagnetic parameters. Substitution of (5-14) and (5-8) into (2-1) gives:

$$2k_0^4 \left(\frac{2\Phi_0}{\tilde{h} c}\right)^4 = \frac{8\pi G}{c^4} k_2^2 \lambda^4 \frac{\Phi_0^2}{2\pi \tilde{h} c} \tag{6-1}$$

so $\quad G = \dfrac{8 k_0^4 \Phi_0^2 c^4}{k_2^2 \lambda^4 (\tilde{h} c)^3} \quad \left(\dfrac{[m^3]}{[kg][s^2]}\right).\tag{6-2}$

If we would know the quantummechanical parameters we would be able to check if the theory as presented is correct. Let us see if it possible to assign meaningful values to these parameters. Let the W-boson, which binds quark and antiquark in the doublet (pi-meson), be characterized by $\tilde{h}\omega = m'_W$. The frequency $\omega$ is determined by the following relationship in quantum mechanical harmonic oscillation:

$$m_{eff} \omega^2 / 2 = \Phi_0 k_2 \lambda^2 .\tag{6-3}$$

Herein is $m_{eff}$ the effective mass of the vibrating doublet. (Note that energetic values of masses are indicated by a "dash" and massive values without). As all vibrating mass $m_{eff}$ eventually is radiated, we have:

$$m'_{eff} = m'_W \tag{6-4}$$

This assumption has been discussed and validated in [1] in the derivation of the lepton algorithm (this algorithm relates the mass ratios of leptons). Moreover we have in state of minimum energy:

$$\lambda d_{min} = d'_{min} = 0,852 \tag{6-5}$$

After substitution of (6-3)-(6-5) into (6-2) we get:

$$G = \frac{1}{2} \frac{k_0^4 (\tilde{h} c) c^4}{k_2^4 \, d'^8_{min} \, m'^2_W} \tag{6-6}$$

As the bosonic energetic mass figure of W-bosons is known in terms of rest mass a relativistic correction of (6-6) is required. So far the metrics as mass and distances have been considered from the viewpoint of an observer in the center of the doublet. However in the labframe the doublet (pi-meson) is moving with near light velocity. Proper correction requires taking the pi-meson velocity $v_\pi$ into account. So, equation (6-6) has to be corrected into:

$$G = \frac{1}{2}\frac{k_0^4(\tilde{h}c)c^4}{k_2^4 \, d'^{8}_{min}}\frac{1\Delta_\pi}{m'^2_{0W}} \quad \text{with} \quad \Delta_\pi = 1 - (v_\pi/c)^2, \tag{6-7}$$

wherein $m'_{0W}$ is the energy equivalent of the W-boson's rest mass. The only unknown parameter left is the pi-meson velocity $v_\pi$. This parameter can be determined from the measurable half life time of pi-mesons. This can be seen as follows. Let $N(t)$ the total number of particles radiated from a centric source. The number $N(t)$ of particles in a wave front after time $t$ is:
:

$$N(t) = N_0 \exp[-\gamma t], \tag{6-8}$$

wherein $\gamma$ is the decay rate. If $v$ is the velocity of the particles then the number remaining at distance $r$ is:

$$N(t) = N_0 \exp[-(\gamma/v)r]. \tag{6-9}$$

So, the force executed by bosons (force carrying particles) per unit of area is:

$$F(r) \sim \frac{\exp[-(\gamma/v)r]}{r^2}. \tag{6-10}$$

(The symbol "~" expresses proportionality). This exponential force behavior corresponds with the force behavior as a consequence of the inter nucleon potential. This potential has a shape as expressed by $f_i[x, y, z]$ in (5-15). (The potential of a quark as adopted in [1] has been based upon the observation that the spatial format of the Higgs field corresponds with the the shape of the inter nucleon potential). As in the region of the attractive force:

$$V(r) \approx \Phi_0 \left( \frac{\exp[-\lambda r]}{\lambda r} \right), \tag{6-11}$$

this force behavior is:

$$F(r) \sim \frac{\exp[-\lambda r]}{r^2}. \tag{6-12}$$

This gives the following relationship:

$$\lambda = \gamma/v. \tag{6-13}$$

The decay parameter $\gamma$ can be related with with the half life of the particles. Half life is the time interval after which the number of particles in the wave front is reduced by a factor 2. As this deals with the wave front, this time interval has to do with the observation of a co-moving observer. The time frame of analysis of half time therefore is proper time $\tau$ rather than (stationary) observable time $t$. Let proper time half life be $\tau_0$.

So, we have:

$$N(\tau + \tau_0) = N_0 \exp[-\gamma(\tau + \tau_0)] = N(\tau)/2 . \tag{6-14}$$

Therefore:
$$\gamma \tau_0 = \log[2] . \tag{6-15}$$

Let $v_\pi$ be the velocity of pi-mesons. Then half time $t_0$ in observable time is:

$$t_0 = \frac{\tau_0}{\sqrt{1 - (v_\pi/c)^2}} = \frac{\log[2]}{v_\pi \lambda_\pi \sqrt{1 - (v_\pi/c)^2}} . \tag{6-16}$$

As the inter nucleon potential has the same shape as the hypothesized quark potential [1] we have from (6-5):

$$\lambda_\pi d_{min} = d'_{min} = 0,852 \tag{6-17}$$

The distance $d_{min}$ corresponds with a quarter of wave length of the pi-meson. This wave length can be found from:

$$\tilde{h}\omega_\pi = m_{0\pi} c^2 \tag{6-18}$$

so that
$$\lambda_\pi = \frac{4}{\pi} \frac{d'_{min} m'_{0\pi}}{\tilde{h}c} \tag{6-19}$$

Under consideration of (6-19) and (6-16) it is found that:

$$\sqrt{\Delta_\pi} = \frac{\pi \log[2]}{4 d'_{min} c t_{0\pi} m'_{0\pi}} (\tilde{h}c) \tag{6-20}$$

In summary: We have calculated the gravitational constant in terms of quantum mechanical parameters. The full expression is;

$$G = \frac{1}{2}\frac{k_0^4(\tilde{h}c)c^4}{k_2^4\, d'^8_{min}}\frac{1\Delta_\pi}{m'^2_{0W}} \quad \text{with} \quad \sqrt{\Delta_\pi} = \frac{\pi\log[2]}{4d'_{min}ct_{0\pi}}\frac{(\tilde{h}c)}{m'_{0\pi}} \qquad (6\text{-}21)$$

The physical quantities needed to calculate the gravitational constant are:

1. Planck's constant and the vacuum light velocity
2. The rest mass of the pi-meson
3. Half life of pi-mesons
4. The rest mass of the charged W-boson

Furthermore four dimensionless constants are required. These constants can be calculated with precision from the spatial Higgs field format. As shown in table I the calculated gravitational constant ($G \approx 7,14 \times 10^{-11}$ m³kg⁻¹s⁻²) shows a surprisingly good fit with the known value $G \approx 6,67 \times 10^{-11}$ . m³kg⁻¹s⁻² .

**Table 14:**

| physics | this theory | calculated |
|---|---|---|
| $\tilde{h}c \approx 197,3$ MeV fm | $k_0 = -0,5$ | |
| $m'_{0\pi} = 139,6$ MeV/c² | $k_2 = 2,37$ | $\sqrt{\Delta_\pi} = 1,15 \times 10^{-16}$ |
| $t_{0\pi} = 2,6 \times 10^{-8}$ s | $d'_{min} = 0,852$ | $G \approx 7,14 \times 10^{-11}$ m³kg⁻¹s⁻² |
| $m'_{0W} = 80,4$ GeV/c² | | |
| | | |